\documentclass[fleqn,10pt]{wlscirep}
\title{An interpolatory ansatz captures the physics of one-dimensional confined Fermi systems}

\author[1]{M.~E.~S. Andersen}
\author[1]{A.~S. Dehkharghani}
\author[2,1]{A.~G. Volosniev}
\author[3,4]{E.~J. Lindgren}
\author[1,*]{N.~T. Zinner}

\affil[1]{Department of Physics and Astronomy, Aarhus University, DK-8000 Aarhus C, Denmark}
\affil[2]{Institut f{\"u}r Kernphysik, Technische Universit{\"a}t Darmstadt, 64289 Darmstadt, Germany}
\affil[3]{Theoretische Natuurkunde, Vrije Universiteit Brussel, and International Solvay Institutes, Pleinlaan 2, B-1050 Brussels, Belgium}
\affil[4]{Physique Th\'eorique et Math\'ematique, Universit\'e Libre de Bruxelles, Campus Plaine C.P.\ 231, B-1050 Bruxelles, Belgium}

\affil[*]{zinner@phys.au.dk}

\begin{abstract}
Interacting one-dimensional quantum systems play a pivotal role in physics.
Exact solutions can be obtained for the homogeneous case using the Bethe ansatz and bosonisation techniques.
However, these approaches are not applicable when external confinement is present.
Recent theoretical advances beyond the Bethe ansatz and bosonisation allow us to predict the behaviour of one-dimensional confined systems with strong short-range interactions, and new experiments with cold atomic Fermi gases have already confirmed these theories.
Here we demonstrate
that a simple linear combination of the strongly interacting solution 
with the well-known solution in the limit of vanishing interactions provides a simple and 
accurate description of the system for 
all values of the interaction strength. This indicates that one can 
indeed capture the physics of confined one-dimensional systems by knowledge of the
limits using wave functions that are much easier to handle than the output of typical 
numerical approaches. We demonstrate our scheme for experimentally relevant systems with 
up to six particles. Moreover, we show that our method works also in the case of mixed systems of particles with different masses.
This is an important feature because these systems
are known to be non-integrable and thus not solvable by the Bethe ansatz
technique.
\end{abstract}
\begin{document}

\flushbottom
\maketitle

\thispagestyle{empty}


\section*{Introduction}
Understanding the properties of low-dimensional systems is not merely an 
academic pursuit. Technologically promising systems such as nanotubes, 
nanowires, and organic conductors have one-dimensional 
nature \cite{giamarchi2003,bezryadin2013,altomare2013}, while 
there is much evidence that high-temperature superconductors owe their
spectacular properties to an effective two-dimensional structure \cite{anderson1987,lee2006}.
However, in the case of interacting particles in one dimension there are still 
many outstanding fundamental issues in their quantum mechanical description. 
An avenue within which these problems can be studied is that of cold atomic gases \cite{lewenstein2007,bloch2008}
where experiments in one-dimensional (1D) confinement can be performed with tunable
interactions for systems of bosons 
\cite{moritz2003,stoferle2004,kinoshita2004,paredes2004,kinoshita2006,haller2009,haller2010} 
or fermions \cite{pagano2014}. Most recently, 1D Fermi systems have been constructed 
with full control over the particle number \cite{serwane2011} and thus engineered 
few-body systems are now available. This provides opportunities to study pairing, 
impurity physics, magnetism, and strongly interacting particles from the
bottom up \cite{zurn2012,zurn2013,wenz2013,murmann2015a,murmann2015b}.

The role of strong interactions in important quantum phenomena such as 
superconductivity and magnetism drives research into the regime of strong interactions
also for 1D systems. In this respect, new theoretical approaches to confined quantum systems with 
strong short-range interactions have been proposed in the last couple of years
\cite{volosniev2014,cui2014,deuret2014,volosniev2015,levinsen2015,yangpu2015,hu2015,yangcui2015}. Within these 
new developments it has become clear that the strongly interacting limit 
has an emergent Heisenberg spin model description. While this was realized 
some time ago for the case of a homogeneous system \cite{ogata1990}, in the 
presence of confinement the Heisenberg model obtained has non-trivial 
nearest-neighbour interactions that depend on the confining potential. This 
can be exploited for tailoring systems to have desired static and dynamic 
properties \cite{volosniev2015}. In the opposite limit where we have vanishing interactions, confined 1D 
systems are trivially solved as the single-particle Schr{\"o}dinger equation 
can be solved numerically to any desired level of accuracy. The natural 
question to ask is how to describe 1D systems with interaction strengths 
that are somewhere in the region between the two extreme limits? 

In this paper, we propose a deceptively simple ansatz that linearly combines our 
knowledge of the weakly and strongly interacting limits. As an ansatz
containing just two wave functions it is much easier to work with as compared
to entirely numerical approaches, which typically have the solution 
represented on a large basis set with many non-zero contributions. 

To describe the main idea we will focus on two-component Fermi systems
of $N_\uparrow$ particles with spin projection up and $N_\downarrow$ particles with 
spin projection down. 
The Hamiltonian for the $N=N_\uparrow+N_\downarrow$ system
is
\begin{equation}\label{ham}
H=\sum_i\left[\frac{p_{i}^{2}}{2m}+V_\mathrm{ext}(x_i)\right]+g\sum_{i<j}\delta(x_i-x_j),
\end{equation}
where $m$ is the mass and the sums run over the number of particles, $i=1,\ldots,N$, 
$x_i$ is the coordinate, and $p_i$ the momentum of the $i$th particle.
The external confinement, $V_\mathrm{ext}(x)$, is assumed to be the same for all particles.
Furthermore, we will assume that $V_\mathrm{ext}(x)$ is parity invariant and has at least $N$ bound states.
The interaction strength $g$ is positive for repulsive interactions and negative for attractive interactions.
We note that since the parity operator commutes with the Hamiltonian, parity is conserved when $g$ changes.
While we only discuss the two-component Fermi system below, 
the case of bosons or Bose-Fermi mixtures is similar in spirit and in 
formalism and we return briefly to this extension in the outlook.
For the majority of the discussion we assume all particles have equal mass, $m$, 
but we also discuss the important extension to mass-imbalanced systems.
Notice
that in the case of two-component Fermi systems, 
particles of the same spin projection will not interact due to the antisymmetry 
required by the Pauli principle which implies that the zero-range interaction 
will vanish for similar spin projections. For this reason we may use 
the general form of the Hamiltonian in Eq.~\eqref{ham} with the sum over $i<j$ 
also for fermionic systems.

Consider now an $N$-body system 
and assume that we have solved the problem for $g=0$ with energy eigenstate
$|\gamma_0\rangle$ and for $1/g=0$ with eigenstate $|\gamma_\infty\rangle$. 
Now form the linear combination
\begin{equation}\label{psi}
|\gamma\rangle=\alpha_0 |\gamma_0\rangle+ \alpha_\infty |\gamma_\infty\rangle.
\end{equation}
This interpolatory ansatz is motivated by the intuitive idea that the wave function of a system with intermediate-strength interactions contains a mixture of qualities from the wave functions with weak and strong interactions.
We may compute $\langle{\gamma|H|\gamma}\rangle$ as a function of $\alpha_0$ and 
$\alpha_\infty$ and look for optimum values. As we shall demonstrate in this paper,
$|\gamma\rangle$ provides a simple yet accurate description of the system for 
any value of $g$. The only exception is deeply bound states to which we return below. 

We will demonstrate that the interpolatory ansatz in Eq.~\eqref{psi}
can capture the qualitative features of the eigenstates of the Hamiltonian 
in Eq.~\eqref{ham} and is quantitatively accurate at the level of a few percent.
Furthermore, we show that the expression for the optimum energy of $|{\gamma}\rangle$ may be modified slightly to make it perturbatively correct in both limits of the interaction strength.
With this modification, the interpolatory ansatz provides an approximation to the eigenenergy with an accuracy that is comparable to state-of-the-art numerical methods, though the ansatz is far simpler.

We provide a proof of principle by considering some important examples from the few-body limit
that are experimentally relevant at the moment. For this purpose we restrict
to a harmonic confinement, i.e. $V_\mathrm{ext}(x)=\tfrac{1}{2}m\omega^2x^2$, throughout
(here $\omega$ is the angular frequency of the oscillator).

The results we present here
are
\begin{itemize}
\item Analytical expressions for the ansatz parameters $\alpha_0$ and
$\alpha_\infty$ that depend exclusively on two matrix elements in $|{\gamma_0}\rangle$ and $|{\gamma_\infty}\rangle$ (as well as the eigenenergies of these limiting states).
\item The case $N=2$ where one has the exact solution available \cite{busch1998}.
Here we show that our method is accurate for both repulsive and attractive 
interactions to less than a few percent. 
\item The $N=3$ case where no analytical solutions are known for general $g$. Here
our ansatz provides very accurate results for all $g$. The results are compared to exact
numerical diagonalisation utilising a unitary transformation of the interaction Hamiltonian.
\item Energies for the impurity limit with $N_\downarrow=1$ and 
$N_\uparrow=1-5$ which we compare to experiments and find excellent 
agreement.
We also discuss the Anderson orthogonality catastrophe for this system,
which is related to the coefficient $\alpha_0$.
\item An extension of the method to systems with particles of 
different mass. The examples we discuss are three-body systems and we compare
to exact numerical results based on the correlated Gaussian method. This is 
the first application of the correlated Gaussian method to mass-imbalanced systems in 1D
that we are aware of.
\end{itemize}

The ansatz we propose can be used to get very simple expressions for different
observables as one needs to compute only a few matrix elements between 
the $g=0$ and $1/g=0$ states of interest. Our method is directly extendable to
bosonic systems or mixed systems, as long as one has access to the two limiting 
wave functions, and while we have focused on harmonically confined systems it can 
be straightforwardly extended to any other form of confinement. 
Furthermore, one may systematically improve the ansatz by
adding more states.

\section*{Results and Discussion}
Let $|{\gamma_0}\rangle$ be an energy eigenstate in the non-interaction limit (that is, $g=0$) with eigenenergy $E_0$. 
We now adiabatically change the interaction strength from $g=0$ to $|g|\to \infty$ and in turn we 
adiabatically change $|{\gamma_0}\rangle$ into a new state denoted $|{\gamma_\infty}\rangle$ with energy $E_\infty$. 
The wave function of the state $|{\gamma_\infty}\rangle$ vanishes whenever the position coordinates of any two particles coincide, and $|{\gamma_\infty}\rangle$ is thus unaffected by the interaction potential
\begin{equation}
V = g \sum_{i<j} \delta(x_i - x_j).
\end{equation}
As our ansatz we now construct the trial state
\begin{equation}
|{\gamma}\rangle = \alpha_0 |{\gamma_0}\rangle + \alpha_\infty |{\gamma_\infty}\rangle, \label{eq:method:trial-state}
\end{equation}
where $\alpha_0$ and $\alpha_\infty$ are real parameters.
Assuming $|{\gamma_0}\rangle$ and $|{\gamma_\infty}\rangle$ are normalised, the energy of the trial state is
\begin{equation}
E = \frac{\langle{\gamma|H|\gamma}\rangle}{\langle{\gamma|\gamma}\rangle} = E_0 + \frac{\langle{\gamma_0|V|\gamma_0}\rangle\alpha_0^2 + \Delta E\alpha_\infty^2}{\alpha_0^2 + \alpha_\infty^2 + 2\langle{\gamma_0|\gamma_\infty}\rangle\alpha_0\alpha_\infty}, \label{eq:method:trial-state-energy}
\end{equation}
where we let $\Delta E \equiv E_\infty - E_0$.

We use a variational approach to solving the Schr{\"o}dinger equation by identifying 
stationary points of the trial state energy functional Eq. \eqref{eq:method:trial-state-energy}. 
We thus select the values of $\alpha_0$ and $\alpha_\infty$ that optimise the energy of the trial state for a given value of $g$. 
This will yield energies and eigenstates that, although approximate, turn out to be extremely accurate as discussed below.

Before presenting the results of the variational calculation, we shall briefly examine Eq. \eqref{eq:method:trial-state-energy} in the limiting cases of the interaction strength:
If we require that $\alpha_\infty/\alpha_0\to 0$ for $g \to 0$ such that the trial state approaches $|{\gamma_0}\rangle$ in the non-interacting limit, Eq. \eqref{eq:method:trial-state-energy} gives a first-order term in $g$ of the form $\langle{\gamma_0|V|\gamma_0}\rangle$, in agreement with first order perturbation theory. We caution, however, that requiring $\alpha_0/\alpha_\infty \to 0$ (i.e. $|{\gamma}\rangle \to |{\gamma_\infty}\rangle$) for $1/g \to 0$, does not automatically ensure that the first order expansion of Eq. \eqref{eq:method:trial-state-energy} in $1/g$ is equal to that of the exact eigenstate. We will come back to this point later on.

The coefficients that yield stationary points of Eq. \eqref{eq:method:trial-state-energy} are given by (see the Supplementary Materials for details)
\begin{equation}
\left(\frac{\alpha_0}{\alpha_\infty}\right)_\mathrm{opt}^{(\pm)} = \frac{\Delta E -\langle{\gamma_0|V|\gamma_0}\rangle
\mp\sqrt{\left(\Delta E -\langle{\gamma_0|V|\gamma_0}\rangle\right)^2+4\langle{\gamma_0|V|\gamma_0}\rangle\Delta E\langle{\gamma_0|\gamma_\infty}\rangle^2}}{2\langle{\gamma_0|V|\gamma_0}\rangle\langle{\gamma_0|\gamma_\infty}\rangle}. \label{eq:method:extreme-trial-state-coefficients}
\end{equation}
This gives the energy
\begin{equation}
E_\mathrm{opt}^{(\pm)} = E_0 +\frac{\langle{\gamma_0|V|\gamma_0}\rangle + \Delta E
\pm \sqrt{(\langle{\gamma_0|V|\gamma_0}\rangle + \Delta E)^2 - 4\langle{\gamma_0|V|\gamma_0}\rangle\Delta E\left(1-\langle{\gamma_0|\gamma_\infty}\rangle^2\right)}}{2\left(1-\langle{\gamma_0|\gamma_\infty}\rangle^2\right)}. \label{eq:method:extreme-trial-state-energy}
\end{equation}
The denominator in Eq. \eqref{eq:method:extreme-trial-state-energy} is positive because $|\langle{\gamma_0|\gamma_\infty}\rangle|\leq 1$ by the Cauchy-Schwarz inequality. Hence, $E_\mathrm{opt}^{(+)}$ is the energy maximum while $E_\mathrm{opt}^{(-)}$ is the energy minimum. To ensure the correct energies in the limits $g=0$ and $1/g=0$, we use the energy minimum, $E_\mathrm{opt}^{(-)}$, to approximate the eigenenergy whenever $g>0$, while for $g<0$ we use the energy maximum, $E_\mathrm{opt}^{(+)}$.

While the interpolatory ansatz is extremely simple, it has a shortcoming in the $1/g\to 0$ limit where it
does not reproduce the slope of the energy. 
As is shown in the following, we may, however, modify the ansatz slightly to correct this behaviour.
Letting $q\equiv -1/g$, the first-order expansion $E_\mathrm{opt} = E_\infty + K_\mathrm{opt}^\infty q$ has the slope
\begin{equation}
K_\mathrm{opt}^\infty = \frac{\partial E_\mathrm{opt}}{\partial q}\bigg|_{q=0} = \frac{\Delta E^2}{K^0} \langle{\gamma_0|\gamma_\infty}\rangle^2, \label{eq:method:strong-interactions-slope}
\end{equation}
where $K^0 = \langle{\gamma_0|V|\gamma_0}\rangle/g$ is the corresponding slope of the energy curve in the limit of vanishing interactions.
This demonstrates that the important quantity in the slope is the overlap $\langle{\gamma_0|\gamma_\infty}\rangle$. 
In the original philosophy of the ansatz, we exploit that we know both wave functions in this overlap exactly and thus
also the overlap itself; this leaves no unfixed parameters. Realising that this yields a discrepancy we have explored
how to modify this assumption in order to improve the approximation.

To this end, we note that the derivation of Eq. \eqref{eq:method:extreme-trial-state-energy} actually does not depend on $|{\gamma_\infty}\rangle$ being an energy eigenstate. It must be a state with energy $E_\infty$ (with respect to the non-interacting Hamiltonian), but 
beyond that the only requirement of $|{\gamma_\infty}\rangle$ is that $V|{\gamma_\infty}\rangle=0$.
Furthermore, $E_\mathrm{opt}$ only depends on $|{\gamma_\infty}\rangle$ through the squared wave-function overlap $\langle{\gamma_0|\gamma_\infty}\rangle^2$. 
Thus, if we substitute $\langle{\gamma_0|\gamma_\infty}\rangle^2$ in Eqs. \eqref{eq:method:extreme-trial-state-energy}--\eqref{eq:method:strong-interactions-slope} with some parameter $\lambda$, we may regard $E_\mathrm{opt}$ and $K_\mathrm{opt}^\infty$ as functions of $\lambda$. 
We can then select $\lambda$ such that $E_\mathrm{opt}(\lambda)$ becomes perturbatively correct, that is $K_\mathrm{opt}^\infty(\lambda) = K_\mathrm{exact}^\infty$, or,
\begin{equation}
\lambda = \frac{K^0 K_\mathrm{exact}^\infty}{\Delta E^2}, \label{eq:method:modified-ansatz-lambda}
\end{equation}
where $K_\mathrm{exact}^\infty$ is the slope of the true eigenenergy curve at $q=0$ 
(which is known exactly using the formalism of A.G.V. {\it et al.}\cite{volosniev2014}).
We shall refer to this perturbatively correct modification, $E_\mathrm{opt}(\lambda)$, as the \emph{modified ansatz}.
Note that this modification breaks variational bounds since we have no \emph{a priori} knowledge of any trial state whose energy is $E_\mathrm{opt}(\lambda)$.
By going from the original interpolatory ansatz to the modified ansatz, we lose information about the wave function, but gain the correct slope of the energy at $1/g \to 0$.
Below we will see that the modified ansatz increases the accuracy significantly as compared to the original ansatz.

We now proceed to discuss examples of the ansatz. Throughout our discussion, 
the external potential is taken to be harmonic 
$V_\textrm{ext}(x)=\tfrac{1}{2}\omega^2m x_{}^{2}$ and the same for 
all particles. This is the most widely studied and experimentally relevant case at
the moment so it will be our focus here. 
Henceforth, we use natural units $\hbar=m=\omega=1$ such that energies
are given in units of $\hbar\omega$, lengths in units of $\sigma=\sqrt{\hbar/m\omega}$, and
interaction strengths $g$ in units of $\sigma\hbar\omega$.

\subsection*{Two particles}
The $N=2$ case is special as analytical results for any $g$ are available due
to the seminal work of Busch {\it et al.} \cite{busch1998}. It is therefore an important benchmark case for our 
approach. First we note that we 
will only be interested in the relative energy as the center of mass 
decouples in the harmonic trap. Note that this decoupling is not an essential assumption 
of our method and is merely a convenience.

\begin{figure}[p]
\begin{center}
\includegraphics{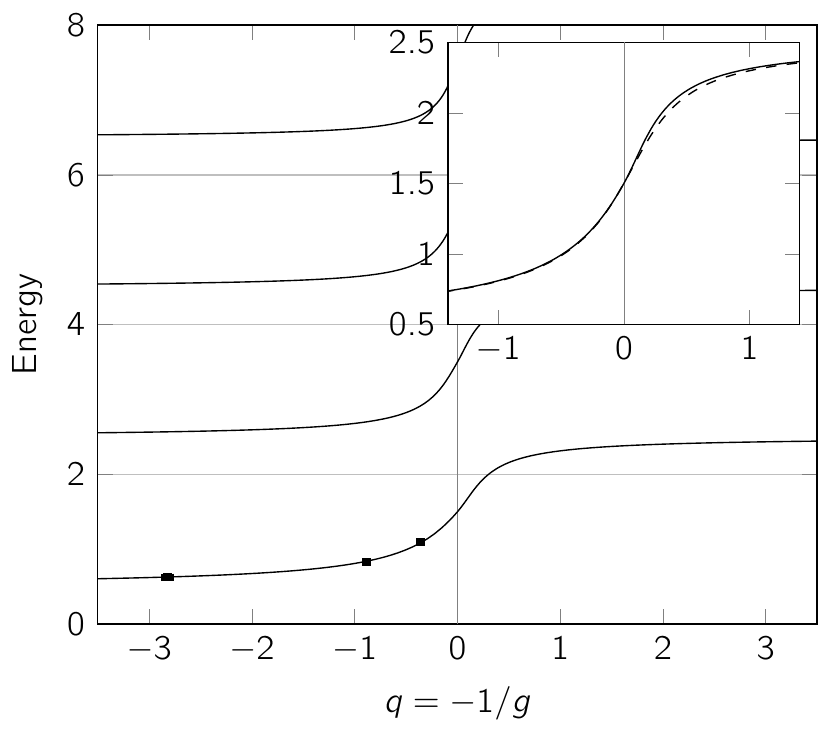}
\caption{{\bf Energy spectrum for the two-particle system.}  Relative energy of two distinguishable fermions according to the interpolatory ansatz, showing the low-energy part of the spectrum. Only states with even parity are plotted as odd-parity states are not influenced by the zero-range interaction. The system also has a deeply bound state, but this is not plotted. Squares are experimental data points (note that the error bars are smaller than the data points) \cite{wenz2013}. The inset is a zoom of the ground-state energy comparing the interpolatory ansatz (solid) with the exact eigenenergy (dashed) for the ground state on the repulsive side as it crosses to the attractive side.}
\label{fig:two-particles-energy-spectrum}
\end{center}
\end{figure}

The details on how to construct the ansatz states for the two-particle case are
given in the Methods section below. 
The energy spectrum using the interpolatory ansatz of 
Eq. \eqref{eq:method:extreme-trial-state-energy} is shown in Fig.\ \ref{fig:two-particles-energy-spectrum}.
Only the even-parity solutions are shown on the plot as odd-parity states are 
unchanged by the zero-range interaction. 
The figure also includes experimental measurements of the ground-state energy.
The experiment has been conducted by S.~Jochim's ultra-cold atoms group in 
Heidelberg, and the data originates from Wenz {\it et al.}\cite{wenz2013}. This data has been 
corrected for the imperfections of the trap as described in Wenz {\it et al.}\cite{wenz2013}.
As seen, the experimental data agrees with the interpolatory ansatz well within the 
experimental uncertainties.

If we expand the energy in terms of $g$, the first-order term agrees with the result of ordinary non-degenerate perturbation theory.
Similarly, in the limit $1/g=0$, the energy curve (as a function of $q$) of the true eigenstate has the same slope as that of our interpolatory ansatz.

The inset in Fig.\ \ref{fig:two-particles-energy-spectrum} shows a zoom of the energy spectrum on the ground state and compares the energy predicted by Eq. \eqref{eq:method:extreme-trial-state-energy} with the exact solution.
For $q>0$ the energy of the interpolatory ansatz is within $0.05$ of the exact energy in the vicinity of $q\sim 0.4$ and even less elsewhere. The error decreases as we move up in the spectrum.
For $q<0$ the deviation is less than $0.006$ for the ground state; again greatest around $q\sim -0.4$. On this side, the error increases for the excited states, but we find that it is bounded by about $0.03$.

We conclude that the ansatz is extremely accurate for the $N=2$ case where we can compare
to analytical results \cite{busch1998}.
One may also compare the wave functions and again find extremely good agreement (see Supplementary Fig. S1 online).

For attractive interactions, a deeply bound molecular state exists that we have so far ignored.
However, it turns out that the ansatz of 
Eq.~\eqref{eq:method:extreme-trial-state-energy} can be extended to also give extremely accurate
results for the deeply bound state. As is shown in the Supplementary Materials, this can be done by including an additional state in the 
ansatz that has the correct asymptotic behaviour as $g\to -\infty$.
We stress that this is in fact in complete agreement with 
the universal philosophy of the ansatz method, i.e. interpolation between (known) extremes.
Thus to address deeply bound states one needs a state in the extreme limit of large negative
energy. This yields a very precise approximation 
also for the deeply bound state. This highlights the universal nature of our approach. We 
will not pursue the deeply bound states any further in this paper.

\subsection*{Three particles}
The simplest non-trivial example of a three-body two-component Fermi system has 
$N_\uparrow = 2$ and $N_\downarrow = 1$. The interaction potential is
\begin{equation}
  V = g (\delta(x_1 - x_2) + \delta(x_1 - x_3) + \delta(x_2 - x_3)),
\end{equation}
and we let $x_1$ be the position of the spin-down fermion, while $x_2$ and $x_3$
denote the position of spin-up fermions. Again the third interaction term 
will vanish for identical fermions but we keep it for generality. 

Eigenstates of the harmonic Hamiltonian are described by two quantum numbers, $\nu \ge 0$ and $\mu \ge 1$, which we shall call the radial quantum number and the angular quantum number, respectively. The eigenenergies are
\begin{equation}
E_{\nu,\mu} = 2\nu + \mu + 1. \label{eq:three-particles:harmonic-energy}
\end{equation}
The quantum numbers $\nu$ and $\mu$ can be used to describe the energy eigenstates in both the non-interacting limit and the infinite-interaction limit.

\subsubsection*{Constructing the ansatz}
Recall that 
$|{\gamma_0}\rangle$ and $|{\gamma_\infty}\rangle$ denote eigenstates in the non-interacting limit and the infinite-interaction limit, respectively, and that we are looking for states that are adiabatically connected. 
As states with different radial quantum number $\nu$ are orthogonal, we assume that 
the adiabatically connected states have the same $\nu$ values. Parity $p$ 
is exactly conserved and thus also the same for two states in the ansatz. 
The quantity that changes with $g$ is therefore the $\mu$ quantum number, 
and we call the limiting values $\mu_0$ and $\mu_\infty$, respectively. 
The angular quantum numbers are related by
\begin{equation}
|\Delta \mu| = |\mu_\infty-\mu_0| = \frac{3+(-1)^{\mu_\infty}p}{2}.
\end{equation}
For repulsive interactions, $\mu_0 < \mu_\infty$, and for attractive interactions, $\mu_0 > \mu_\infty$, cf.\ Eq. \eqref{eq:three-particles:harmonic-energy}.
The optimum energy of the trial state can now be determined using Eq. \eqref{eq:method:extreme-trial-state-energy} with $E_0$ and $E_\infty$ given by Eq. \eqref{eq:three-particles:harmonic-energy}.

\subsubsection*{Energy spectrum}

\begin{figure}
\begin{center}
\includegraphics{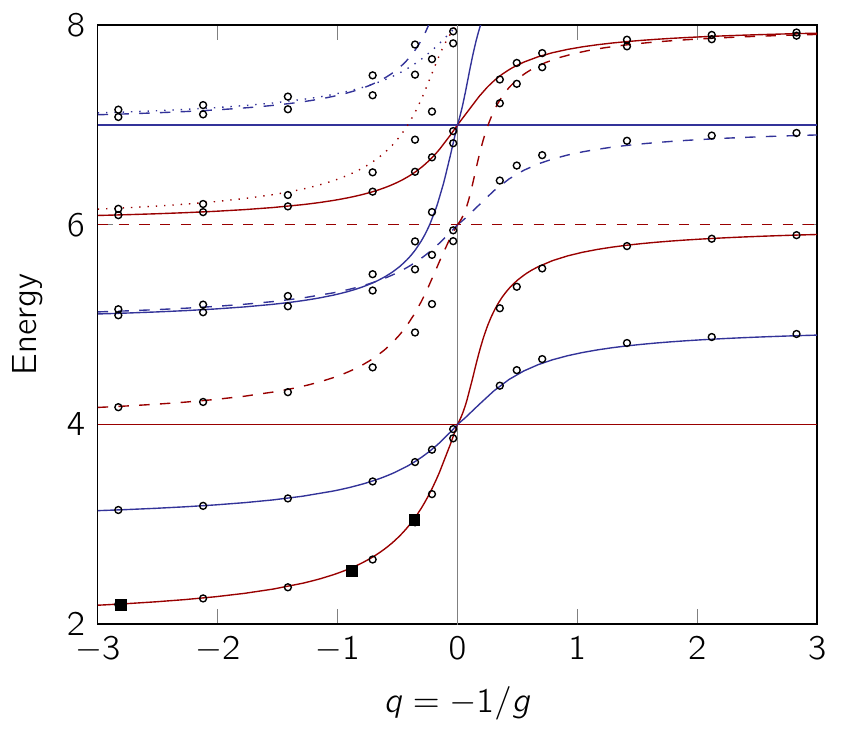}
\caption{{\bf Energy spectrum of the interpolatory ansatz for the three-particle system.}  Red curves are energies of trial states with odd parity, and blue curves are those with even parity. Solid curves represent states with $\nu=0$, dashed curves $\nu=1$, and dotted curves $\nu=2$. Circles are numerical calculations. Deeply bound states are excluded from the plot. Squares are experimental data-points \cite{wenz2013}. The error bars on the experimental data points are smaller than the squares and are therefore not shown.}
\label{fig:three-particles:energy-spectrum}
\end{center}
\end{figure}

Figure \ref{fig:three-particles:energy-spectrum} shows the energy spectrum for $E < 8$ (deeply bound states are not considered) both as calculated using the interpolatory ansatz and by exact numerical diagonalisation.
The interpolatory ansatz clearly describes the qualitative features of the spectrum well; both for repulsive and attractive interactions. In this energy-range, the $\nu=0$ states also give a good quantitative match to the numerical results. However, the error seems to grow with $\nu$.

We see in Fig.\ \ref{fig:three-particles:energy-spectrum} that trial states with $\nu=0$ and $\Delta\mu = 1$ offer a particularly good approximation to the corresponding eigenstates in the repulsive region (see e.g., the first excited state). This is because the slope of the energy curve, that is $K_\mathrm{opt}^\infty = \partial E_\mathrm{opt}^{(-)} / \partial q$, at $q=0$ is the same as that of the exact eigenenergy. For the trial states with $\Delta \mu = 2$, this is generally not the case. This result suggests that $K_\mathrm{opt}^\infty$ is an important quantity to reproduce correctly in an attempt such as the present to describe the physics of our problem through a simple ansatz.
Note also, that the slope of the energy curve has a discontinuity at $1/g=0$, which contradicts the expectation that 
the states go smoothly through this region.

\begin{figure}
\begin{center}
\includegraphics{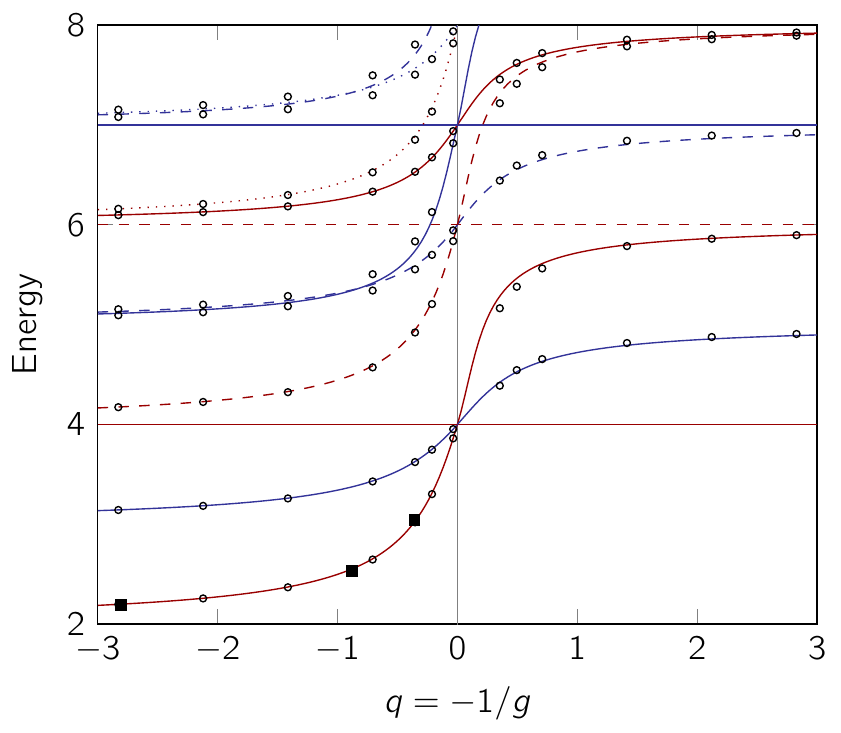}
\caption{{\bf Energy spectrum for the three-particle system -- modified ansatz.}  Energy spectrum for the three-particle system as in Fig.\ \ref{fig:three-particles:energy-spectrum}, but with the interpolatory ansatz modified to be perturbatively correct.}
\label{fig:three-particles:energy-spectrum-exact-slope}
\end{center}
\end{figure}

If we now enforce the correct slope by a modification of the interpolatory ansatz as proposed in the discussion following Eq. \eqref{eq:method:strong-interactions-slope}, we arrive at the spectrum shown in Fig.\ \ref{fig:three-particles:energy-spectrum-exact-slope}.
We see that the modified ansatz agrees better with the numerical results than the original ansatz; especially for states with $\nu>0$.
There are, however, still deviations on the attractive side of the spectrum, and for high energies, also on the repulsive side.
We shall give a quantitative discussion of the quality of the approximation when discussing the impurity system below. We have included the experimental measurements of the ground-state energy \cite{wenz2013} in the figure and
see that both ansatz and modified ansatz agrees very nicely with experiment, although for large $g$ the modified ansatz
naturally does better.

\subsubsection*{Mass-imbalanced systems}
In the case of different masses, one typically uses another length scale given by $\sigma=\sqrt{\hbar/\eta\omega}$ where $\eta=\sqrt{m_1m_2m_3/(m_1+m_2+m_3)}$, and $m_1$ is the mass of the spin-down fermion while $m_2$ and $m_3$ are the masses of the spin-up fermions.
We now consider the case where $m_1=M$ and $m_2=m_3=m$.

\begin{figure}
\begin{center}
\includegraphics{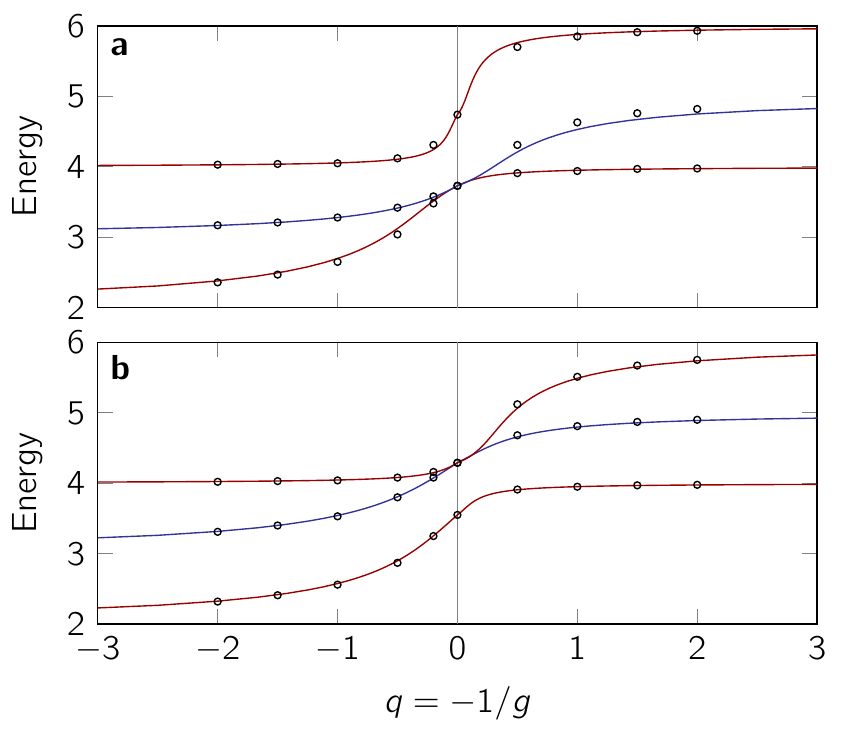}
\caption{{\bf Energy spectrum for the three-particle system with different masses.}  Examples are given with mass ratio ({\bf a}) $M/m=1/2$ and ({\bf b}) $M/m = 2$. Red curves indicate odd-parity states while blue curves indicate even-parity states. The circles are the exact results, calculated numerically using the correlated Gaussian method.}
\label{fig:three-particles:energy-spectrum-unequal-masses}
\end{center}
\end{figure}

In Fig.~\ref{fig:three-particles:energy-spectrum-unequal-masses} we show the energy spectrum obtained by the interpolatory ansatz for $M/m=1/2$ and $M/m=2$ and compare this with numerically calculated results using the correlated Gaussian approach (see the Supplementary Materials for details on the numerical methods). The agreement between the ansatz and the numerical 
results is striking for the low-energy part of the spectrum considered here, and we see that the ansatz can be 
extended also to mass-imbalanced systems.

\subsection*{Impurity systems}
\label{sec:polaron}
We now consider a system of $N$ fermions among which one particle ($x_1$) is spin-down and the $N_\uparrow = N-1$ remaining particles ($x_2,\dotsc,x_N$) are spin-up. Taking into account that interactions 
between identical fermions vanish, we can write the interaction term as
\begin{equation}
V = g \sum_{k=2}^N \delta(x_1 - x_k).
\end{equation}
We restrict the discussion to the ground state with repulsive interactions as  
this has been a focus of recent experimental attention \cite{wenz2013}. The 
considerations can be extended to obtain more states in the spectrum, to
the attractive side, and to deeply bound states in the same manner as in the 
previous examples.

We denote by $|{\gamma_\mathrm{A}}\rangle$ the Slater determinant of the single-particle harmonic 
eigenstates $\langle{x_k|n}\rangle=\psi_{n}(x_k)$ for $k=1,\dotsc,N$ and $n=0,\dotsc,N-1$. 
Here $\psi_n(x)$ is the single-particle eigenstate of the harmonic oscillator Hamiltonian in 
coordinate space with quantum number $n$ and argument $x$.
The state $|\gamma_\mathrm{A}\rangle$ is antisymmetric with respect to interchange of any two coordinates, so $\langle{\gamma_\mathrm{A}|V|\gamma_\mathrm{A}}\rangle=0$. We can write the antisymmetric state as
\begin{equation}
|{\gamma_\mathrm{A}}\rangle = \sum_{k=1}^N |{\gamma_\mathrm{A}}\rangle_k,
\end{equation}
if we define
\begin{equation}
|{\gamma_\mathrm{A}}\rangle_k \equiv \sum_{\substack{\sigma\in S_N \\ \sigma(1)=k}}\; \int\limits_{\Pi(\sigma)} \mathrm{d}{\mathbf{x}} \, |{\mathbf{x}}\rangle\langle{\mathbf{x}|\gamma_\mathrm{A}}\rangle,
\end{equation}
where $\mathbf{x}=(x_1,\dotsc,x_N)$, $S_N$ is the symmetric group of order $N$, and $\Pi(\sigma)$ indicates the integration region $x_{\sigma(1)}<\dotsb<x_{\sigma(N)}$.
In each region, $\Pi(\sigma)$, the wave function of the ground state in the infinite-interaction limit is proportional to $\langle{\mathbf{x}|\gamma_\mathrm{A}}\rangle$. The (normalised) ground state in the infinite-interaction limit is then
\begin{equation}
|{\gamma_\infty}\rangle = \sqrt{\frac{N}{\sum_{i=1}^N a_i^2}} \sum_{k=1}^N a_k |{\gamma_\mathrm{A}}\rangle_k \label{eq:polaron:inf-int-eigenstate}
\end{equation}
for some coefficients $\mathbf{a} = (a_1,\dotsc,a_N)$ to be determined by the method of A.G.V. {\it et al.}\cite{volosniev2014} (see the Supplementary Materials). The ground state in the non-interaction limit is the single-particle harmonic eigenstate $\psi_0(x_1)$ multiplied by the Slater determinant for the remaining particles in the states $\psi_n(x_k)$ for $k=2,\dotsc,N$ and $n=0,\dotsc,N-2$.

\begin{figure}
\centering
\includegraphics{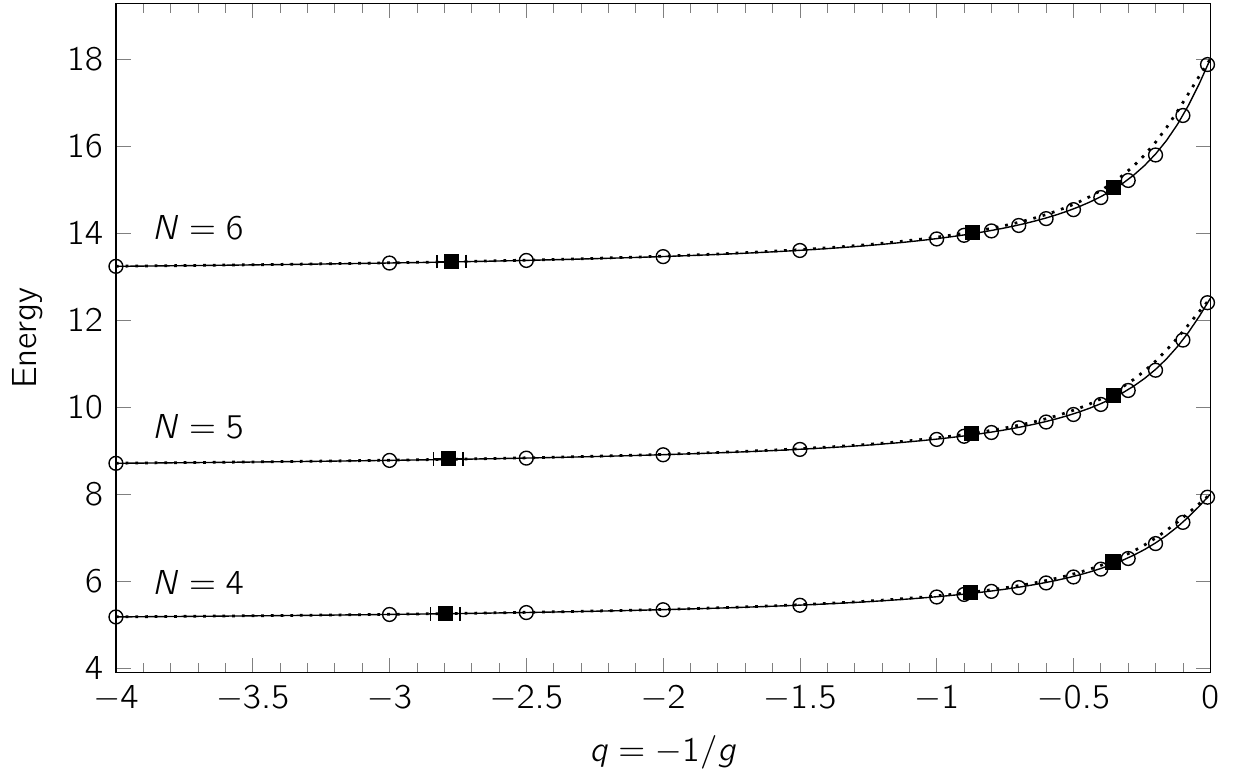}
\caption{{\bf Energy spectrum of the impurity system.}  Energy of the ground state as predicted by the interpolatory ansatz with $|{\gamma_\infty}\rangle$ being the energy eigenstate in the $q=0$ limit (dotted curve) and the modified ansatz (solid curve) for $N=4,5,6$. This is compared with exact numerical calculations (circles) and experimental data (squares) \cite{wenz2013}. Note that the error bars on the experimental data points are smaller than the squares and are only discernible for the points at small $g$.}
\label{fig:polaron:energy-spectrum}
\end{figure}

For systems of four and five particles, the interpolatory ansatz with these $|{\gamma_0}\rangle$ and $|{\gamma_\infty}\rangle$ yields integrals that can be evaluated analytically. For larger systems, the integrals are readily evaluated numerically. The resulting energies for $N=4-6$ are plotted as dashed lines in Fig.\ \ref{fig:polaron:energy-spectrum}. 
Here we see a very good agreement between the ansatz and the numerically exact results, and in turn
excellent agreement with experimental measurements \cite{wenz2013}. This indicates that the energetics
of the system is captured by the interpolatory ansatz with high accuracy.

However, as we discussed briefly above, the ansatz does not generally reproduce the correct first-order energy term at $1/g=0$.
Assuming that the non-interacting state should remain untouched, this prompts us to investigate whether 
another state can be found in the $1/g=0$ limit that can replace $|{\gamma_\infty}\rangle$ in 
Eq. \eqref{eq:method:trial-state} and in turn give the exact result for the slope of the energy at $1/g=0$.
As noted above, Eq.~\eqref{eq:method:extreme-trial-state-energy} remains valid if we substitute the eigenstate $|{\gamma_\infty}\rangle$ with any other state with energy $E_\infty$ that obeys $V|{\gamma_\infty}\rangle=0$.
In particular, any choice of $\mathbf{a}$ in Eq. \eqref{eq:polaron:inf-int-eigenstate} would work, and $E_\mathrm{opt}$ only depends on $\mathbf{a}$ through the wave-function overlap $\langle{\gamma_0|\gamma_\infty}\rangle$. Hence, because $E_\mathrm{opt}$ is monotonic in $\langle{\gamma_0|\gamma_\infty}\rangle^2$, we may optimise the energy with respect to $\mathbf{a}$ for all values of $g$ by maximising $\langle{\gamma_0|\gamma_\infty(\mathbf{a})}\rangle^2$ with respect to $\mathbf{a}$.
Leaving the details to the Supplementary Materials, the optimum is
\begin{equation}
\langle{\gamma_0|\gamma_\infty(\mathbf{a_\mathrm{max}})}\rangle^2 = N \sum_{m=1}^N \langle{\gamma_0|\gamma_\mathrm{A}}\rangle_m^2 \label{eq:optimum-overlap}
\end{equation}
with
\begin{equation}
\mathbf{a_\mathrm{max}} \propto (\langle{\gamma_0|\gamma_\mathrm{A}}\rangle_1,\langle{\gamma_0|\gamma_\mathrm{A}}\rangle_2,\dotsc,\langle{\gamma_0|\gamma_\mathrm{A}}\rangle_N).
\end{equation}
For the ground state of the $N=3-6$ systems, however, $\langle{\gamma_0|\gamma_\infty(\mathbf{a_\mathrm{max}})}\rangle^2$ is very close to the known exact value of $\langle{\gamma_0|\gamma_\infty}\rangle^2$, and is not large enough to make the slope of the energy correct in the strongly-interacting limit, that is, $\langle{\gamma_0|\gamma_\infty(\mathbf{a_\mathrm{max}})}\rangle^2 < \lambda$ with $\lambda$ given by Eq. \eqref{eq:method:modified-ansatz-lambda}.
This indicates that we cannot find a state in the infinite-interaction limit that both has the correct zeroth-order 
energy, $E_\infty$, and satisfies the delta-boundary conditions, $V|{\gamma_\infty}\rangle=0$.
However, we caution 
that this is under the assumptions that the wave function of $|{\gamma_\infty}\rangle$ is continuous and has discontinuous derivatives that 
satisfy the boundary conditions imposed by the zero-range interaction. 

We see that the interpolatory ansatz -- with whatever choice of $|{\gamma_\infty}\rangle$ -- does not reproduce the correct energy slope.
The modified ansatz, however, does.
It is worthwhile to note that the inability to find a state, $|\gamma\rangle=\alpha_0 |\gamma_0\rangle+ \alpha_\infty |\gamma_\infty\rangle$, whose expectation value is that predicted by the modified ansatz, does not imply that the modified ansatz cannot be used.
The two methods rely on different information about the system:
The interpolatory ansatz requires the knowledge of the wave function at $g=0$ and $1/g=0$, whereas the modified ansatz uses the knowledge of the energy behaviour.
Therefore, one can use the modified anzatz to estimate the energies, and the interpolatory ansatz to approximate the wave functions.

\begin{figure}
\centering
\includegraphics{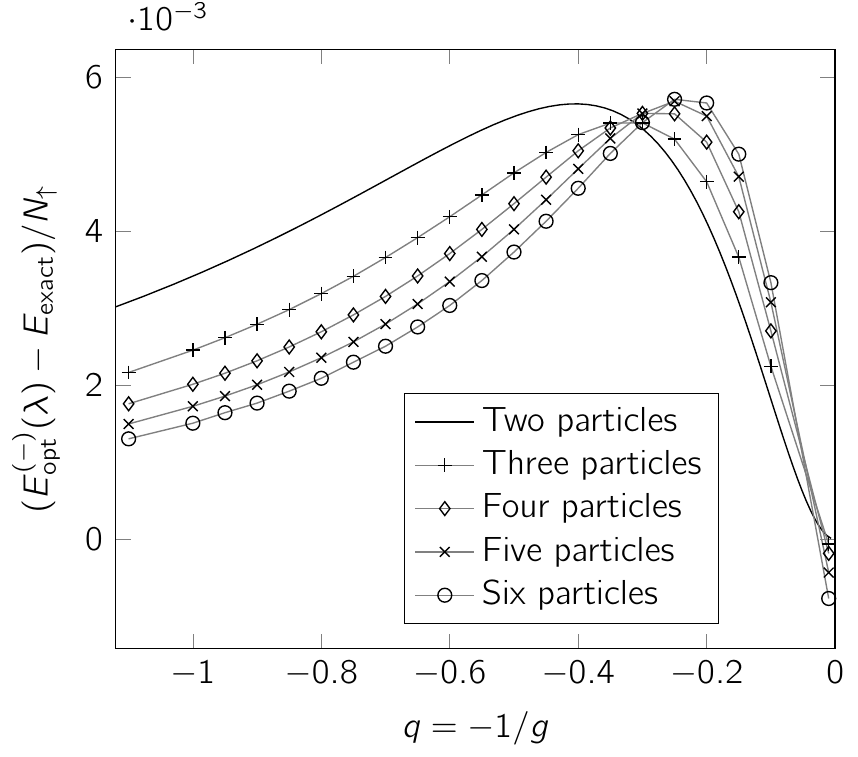}
\caption{{\bf Ansatz accuracy comparison.}  Error in energy according to the modified ansatz compared to exact numerical results for impurity systems of $N=2-6$ particles. (For $N=2$ it is compared to the exact analytical solution.) For $N\ge 3$, the gray lines between the points are a mere guide to the eye. Note the scale of the vertical axis.}
\label{fig:discussion:scaled-energy-error-exact-slope}
\end{figure}

Figure \ref{fig:polaron:energy-spectrum} compares the ground-state energy of the modified ansatz with results from an exact numerical diagonalisation as well as experimental data \cite{wenz2013}. Here we see an improved agreement for large $g$.
The deviations of the modified ansatz from exact numerical diagonalisation are shown in greater detail in 
Figure \ref{fig:discussion:scaled-energy-error-exact-slope}.
The errors are more than an order of magnitude smaller than those of the unmodified ansatz (shown in Supplementary Fig. S4 online).

We plot the error scaled against $N_\uparrow$ instead of $E_0$ or $E$, because $E_0$ scales as $N_\uparrow^2$ while $\Delta E = N_\uparrow$.
As seen in Fig.\ \ref{fig:polaron:energy-spectrum}, the maximum in the error moves to larger interaction strength for higher 
$N_\uparrow$, i.e. with system size. There seems to be only very little increase in the magnitude of the error for larger
$N_\uparrow$ for the system sizes we have studied.
Notice that the error is slightly negative for $1/g\simeq 0$ in the cases $N=3-6$. Since we know the 
exact energies and slopes around $1/g=0$, the likely cause of this is that we are pushing the accuracy of the 
exact numerical diagonalisation method here.

Our first observation is that the approximation of the ground-state energy offered by the modified ansatz is so good that it can compete and even in some cases beat state-of-the-art numerical exact methods. The drawback seems quite 
evident, i.e. we cannot be sure that any state exists in the infinite-interaction limit that when used in Eq. \eqref{eq:method:trial-state-energy} would reproduce the energy of the modified ansatz.
Thus, the modified ansatz 
does not immediately give us any information about the wave function in spite of its near 
perfect approximation of the energy. We also note that the modified ansatz breaks the variational 
bound on the ground state and can in principle have lower energy as compared to the exact result.
However, we have clearly demonstrated that the slope of the energy at $1/g=0$ is an extremely 
important quantity for these systems as it appears crucial to reproduce in order to capture the
energetics. This highlights the important role played by the recently developed 
theory of strongly interacting confined systems \cite{volosniev2014}.

\subsubsection*{Anderson overlap}
Finally, we discuss the so-called Anderson overlap, which is the wave-function overlap between the non-interacting eigenstate, $|{\gamma_0}\rangle$, and the interacting state, $|{\gamma}\rangle$, for some value of the interaction strength, $g$.
This quantity is related to the Anderson orthogonality catastrophe \cite{anderson1967}, which states that the Anderson overlap is zero in the thermodynamic limit; that is $\langle{\gamma_0|\gamma}\rangle\to 0$ for $N\to\infty$, and in particular $\langle{\gamma_0|\gamma_\infty}\rangle\to 0$ for $N\to\infty$. 

\begin{figure}
\begin{center}
\includegraphics{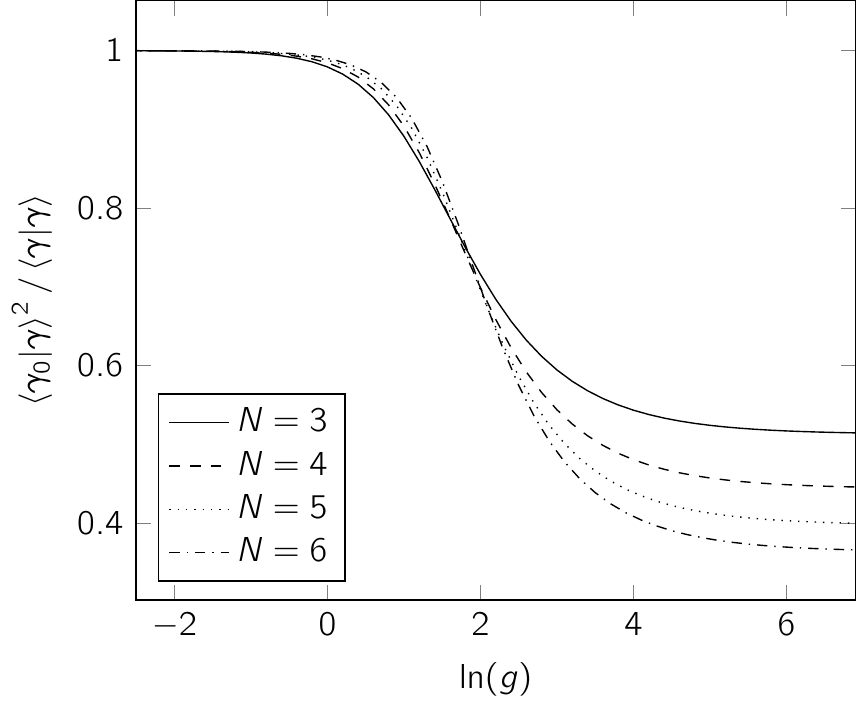}
\caption{{\bf Anderson overlap according to the interpolatory ansatz.}  The (squared) overlap between the non-interacting energy eigenstate and the interacting state as a function of interaction strength for $N=3-6$.}
\label{fig:polaron:anderson-overlap}
\end{center}
\end{figure}

In Fig.~\ref{fig:polaron:anderson-overlap} we illustrate how the overlap $\langle{\gamma_0|\gamma}\rangle^2$ -- with $|{\gamma}\rangle$ given by the interpolatory ansatz -- decreases as a function of $g$. 
Due to the fact that we only consider a finite-sized system, the overlap does not approach zero as $g\to\infty$, but the plot clearly shows that the overlap from our interpolatory ansatz tends to decrease as expected. The fact
that the ansatz gives a very accurate approximation for the energy of the system does not immediately imply that 
this is also the case for the wave function. We leave this question for future studies, and the overlaps presented
here are thus predictions based on the ansatz. It should be compared either to elaborate exact numerical calculations or to 
experimental measurements.

\section*{Outlook}
We have proposed a simple interpolatory ansatz for approximating the energy eigenstate of a confined, one-dimensional system of interacting particles.
The ansatz is a linear combination of known eigenstates in the extreme limits of the interaction strength, $g\to 0$ and $1/g \to 0$, respectively. Thanks to recent advances in the description of the eigenstates in the $1/g=0$ limit, both these wave functions
are now available. An analytical expression for the optimum energy of this ansatz is presented which is an elementary function of only two matrix elements; the interaction energy of the eigenstate at $g=0$, and the wave-function overlap between the eigenstates in the two limits of $g\to 0$ and $1/g\to 0$. By focusing on harmonically trapped impurity systems of fermions, we have demonstrated that the ansatz is able to capture the physics of such a system. 
It gives us a highly accurate approximation for the energy and it also gives us a very simple expression for the wave function.

For both the two- and three-particle systems, we have been able to reproduce the entire energy spectrum with the interpolatory ansatz, save for deeply bound molecular states. The ansatz can be extended to describe deeply bound states as well; this has been shown specifically for the two-particle system.
Taking the three-particle system as an example, we have also demonstrated that the ansatz works equally well for mass-imbalanced systems.
A future extension of this study might investigate mass-imbalanced systems of four or more particles. It should be 
noted that the bottleneck here is that there are generally quite few known results about mass-imbalanced systems 
in the $1/g\to 0$ limit \cite{amin2015}. It is an open problem to find a general method that yields exact eigenstates
in the strongly interacting regime for mass-imbalanced systems.

A drawback of our ansatz is that it is in general not perturbatively correct in the strongly-interacting regime. 
More precisely, if we take the first order derivative of the energy with respect
to $1/g$ it deviates from the known exact result.
We may, however, modify the expression for the energy of the interpolatory ansatz slightly such that it is perturbatively correct to linear order in $g$ for $g\to 0$ and to linear order in $1/g$ for $1/g\to 0$.
The modified ansatz has great simplicity and accuracy at a level that
is competitive with state-of-the-art numerical methods for obtaining the energy
of the ground state for arbitrary $g$. Due to its simplicity it should provide
a very useful tool.

We note that although the results presented here assume that the particles are 
trapped in an external harmonic trap, the formalism is completely general and 
can be applied for arbitrary external potentials with at least $N$ bound single-particle
states for $N$-body systems. As we have shown, the relative deviation of 
the energy obtained from the ansatz grows only very slowly with $N$, and there should
be no problem in extending the technique to even larger systems than considered
here. The decisive quantity is the overlap of the $g=0$ and $1/g=0$ wave functions. 
This overlap may be computed using the same methods that have recently been used
to compute spin chain models for strongly interacting fermions \cite{loft2015b}
and thus scaling to larger $N$ of order 30 or 40 is certainly within reach.

A future direction would be to consider a generalised version of the 
interpolatory ansatz where one systematically adds more states at 
$g=0$ and $1/g=0$ in order to gradually improve the comparison. In addition,
it is relatively straightforward to apply the interpolatory
ansatz in the case of strongly interacting bosons \cite{zinner2013,massignan2015}, 
or mixed systems \cite{volosniev2014,hu2015}. The requirements are knowledge
of states in the two limits and their overlaps so that the interpolation can be performed. 
The formulae for the interpolated energy given here still apply.
An example could be an impurity
interacting strongly with a Tonks-Girardeau gas of hard-core bosons, which 
is a topic of great recent interest \cite{palzer2009,mathy2012,knap2014}.

\section*{Methods}

In the following, we provide the details of the methods used in applying the interpolatory ansatz to two-, three- and many-particle systems.

\subsection*{Details of the two-particle system}
\label{sec:two-particles}

We consider a system of two distinguishable fermions and we define $x = (x_1-x_2)/\sqrt{2}$.
In the absence of interactions, the energy eigenstates are the harmonic eigenstates denoted $|{n}\rangle$ with integer $n\ge 0$ (we are only concerned with the motion relative to the center of mass).

Exact solutions of the two-particle problem are available for arbitrary values of $g$ \cite{busch1998,farrell}.
The energy of an exact energy eigenstate is given indirectly by \cite{farrell}
\begin{equation}
q = \frac{1}{2\sqrt{2}}\frac{\Gamma\big(\frac{1-2E}{4}\big)}{\Gamma\big(\frac{3-2E}{4}\big)}, \label{eq:two-particles:exact-energy}
\end{equation}
and the wave function of the state is
\begin{equation}
\psi(x) = \left( -\sqrt{2}q \; {}_1F_1\!\left(\tfrac{1-2E}{4};\tfrac{1}{2};x^2\right) + |x| \; {}_1F_1\!\left(\tfrac{3-2E}{4};\tfrac{3}{2};x^2\right) \right) e^{-x^2/2}
\end{equation}
where ${}_1F_1$ is the confluent hypergeometric function of the first kind.

\subsubsection*{Constructing the ansatz}
Let $|{\gamma_0}\rangle = |{n_1}\rangle$ be an energy eigenstate in the non-interaction limit.
Here $|{n_1}\rangle$ is a single-particle eigenstate of a harmonic oscillator
Hamiltonian (in the relative coordinate $x$) with quantum number $n_1$. Furthermore, let $|{\gamma_\infty}\rangle$ be the corresponding eigenstate in the infinite-interaction limit -- that is, $|{\gamma_0}\rangle \to |{\gamma_\infty}\rangle$ as the interaction strength is changed adiabatically from $g=0$ to $1/g=0$. By conservation of parity, the states $|{\gamma_0}\rangle$ and $|{\gamma_\infty}\rangle$ have the same parity.

If $|{\gamma_0}\rangle$ is odd, $\langle{x=0|\gamma_0}\rangle = 0$ and thus $\langle{\gamma_0|V|\gamma_0}\rangle = 0$. Hence, odd harmonic eigenstates do not change as we introduce a non-zero interaction. The even harmonic eigenstates do, however, change; the correct even eigenstate in the infinite-interaction limit is
\begin{equation}
|{\gamma_\infty}\rangle = \int\mathrm{d}{x} |{x}\rangle\mathrm{sgn}(x)\langle{x|n_2}\rangle \label{eq:two-particles:even-inf-int-eigenstate}
\end{equation}
with $n_2 = n_1 + 1$ for repulsive interactions ($g>0$) and $n_2 = n_1 - 1$ for attractive interactions ($g<0$).

\subsection*{Details of the three-particle system}
\label{app:three}

Before we employ the interpolatory ansatz, we first separate out the 
center-of-mass motion using hyperspherical coordinates \cite{harshman,zinner2013}. 
This is done merely for convenience and is not in any way essential for
the approach. Defining $x=(x_2-x_3)/\sqrt{2}$ and $y=(x_2+x_3)/\sqrt{6}-2x_1/\sqrt{6}$, the
hyperradius is given by $\rho=\sqrt{x^2+y^2}$ and the hyperangle is defined 
by $\tan\phi=y/x$.
The Hamiltonian of the relative motion becomes
\begin{equation}
H = \frac{1}{2} (\nabla^2 + \rho^2) + V,
\end{equation}
where $\nabla^2$ is the Laplacian in polar coordinates $(\rho,\phi)$ and
\begin{equation}
V = \frac{g}{\sqrt{2}\rho} \sum_{j=1}^6 \delta\!\left(\phi - \frac{2j-1}{6}\pi\right)
\label{samemassinteractionterm}
\end{equation}
is the interaction.

\subsubsection*{Limiting cases}
The quantum numbers $\nu$ and $\mu$ can be used to describe the energy eigenstates in both the non-interacting limit and the infinite-interaction limit. The eigenstate wave function in both limits has the general form \cite{harshman}
\begin{equation}
\psi_{\nu,\mu}(\rho,\phi) = \sqrt{\frac{2\nu!}{(\nu+\mu)!}} L_\nu^\mu(\rho^2) e^{-\rho^2/2} \rho^\mu \Phi_\mu(\phi) \label{eq:three-particles:wave-function},
\end{equation}
where $L_\nu^\mu$ is a generalised Laguerre polynomial.

In the non-interacting limit, the (normalised) angular part of the wave function is
\begin{equation}
\Phi_\mu(\phi) = \frac{1}{\sqrt{\pi}} \cdot \begin{cases}
\cos(\mu\phi), & \text{for } p = -1 \\
\sin(\mu\phi), & \text{for } p = +1,
\end{cases}
\end{equation}
where $p$ is the parity of the wave function. In this limit $p=(-1)^\mu$.

In the infinite-interaction limit, the wave function vanishes at the lines $\phi=-5\pi/6$, $-\pi/2$, $-{\pi}/{6}$, ${\pi}/{6}$, ${\pi}/{2}$, ${5\pi}/{6}$. In the regions between these lines, the wave function solves the Schr\"{o}dinger wave equation for the harmonic oscillator without interactions.

The eigenenergies in the limit $1/g=0$ is given by Eq. \eqref{eq:three-particles:harmonic-energy} with $\mu=3,6,9,12,\dotsc$. Each allowed eigenenergy is three-fold degenerate (not counting states with a different radial quantum number $\nu$): One of the eigenstates in each energy triplet is a harmonic eigenstate with $\mu=3,6,9,12,\dotsc$ which is unaffected by the interactions. Between the two remaining eigenstates in the triplet, one is odd and the other is even.

When $\mu$ is even (that is, $\mu=6,12,18,\dotsc$) the non-trivial eigenstates in the infinite-interaction limit have the angular wave function
\begin{equation}
\Phi_\mu(\phi) = \sqrt{\frac{2-p}{2\pi}} \sin(\mu\phi) \cdot \begin{cases}
-p-1, & \text{for } -\frac{\pi}{6} \le \phi < \frac{\pi}{6} \\
1, & \text{for } \frac{\pi}{6} \le \phi < \frac{\pi}{2},
\end{cases}
\end{equation}
if $p$ is their parity. The coefficients in the four remaining regions follow by symmetry considerations.

When $\mu$ is odd ($\mu=3,9,15,\dotsc$),
\begin{equation}
\Phi_\mu(\phi) = \sqrt{\frac{2+p}{2\pi}} \cos(\mu\phi) \cdot \begin{cases}
p-1, & \text{for } -\frac{\pi}{6} \le \phi < \frac{\pi}{6} \\
1, & \text{for } \frac{\pi}{6} \le \phi < \frac{\pi}{2}.
\end{cases}
\end{equation}

\subsection*{Details of the mass-imbalanced system}
\label{app:details-of-the-mass-imbalanced-system}

The transformation into hyperspherical coordinates 
proceeds along the same lines with a modified Jacobian.
Generally, the coordinates $\mathbf{x}=(x_1,x_2,x_3)$ are transformed to Jacobi coordinates, $\mathbf{x}' = \mathbf{J}\mathbf{x}$, through the transformation matrix
\begin{equation}
\mathbf{J} = \frac{1}{\sqrt{\mu}} \begin{bmatrix}
0 & \mu_{23} & -\mu_{23} \\
-\frac{\mu}{\mu_{23}} & \frac{\mu m_2}{\mu_{23}M_{23}} & \frac{\mu m_3}{\mu_{23}M_{23}} \\
\frac{m_1}{\sqrt{M_{123}}} & \frac{m_2}{\sqrt{M_{123}}} & \frac{m_3}{\sqrt{M_{123}}}
\end{bmatrix},
\end{equation}
where $M_{23}=m_2+m_3$, $M_{123}=m_1+m_2+m_3$ and the `reduced' masses are defined as $\mu_{23}=\sqrt{m_2m_3/M_{23}}$ and $\mu=\sqrt{m_1m_2m_3/M_{123}}$.
This transformation allows us to separate the center-of-mass motion from the relative motion, the solutions of the former being the well-known harmonic eigenstates.
Afterwards, we transform the remaining relative coordinates into hyperspherical coordinates, $\rho$ and $\phi$, by $\rho=\sqrt{x_1'^2+x_2'^2}$ and $\tan(\phi)={x_2'}/{x_1'}$.

From now on, we assume that $m_1=M$ and $m_1=m_2=m$.
The interaction potential can then be written as
\begin{equation}
V=\frac{g}{\rho}\sqrt{\frac{2\zeta}{\zeta^2+1}}\Big(\sum_\pm \delta\big(\phi\pm\theta_0\big)+\delta\big(\phi\pm\theta_0-\pi\big)\Big), \label{eq:three:unequal-masses-V}
\end{equation}
where $\theta_0=\arctan(\zeta)$ and $\zeta\equiv\mu/m=\sqrt{M/({2m+M})}$.
As for the equal-mass case, the energy is given by Eq. \eqref{eq:three-particles:harmonic-energy} \cite{loft2015a}.

The $\mu$ eigenvalue can be found by using parity symmetry, the Pauli principle and the delta-boundary conditions of the interaction potential.
Using these conditions, one can setup an equation from which $\mu$ can be obtained in the limits $g=0$ and $1/g=0$:
For solutions with odd parity, $\mu$ solves the equation
\begin{equation}
\cos(\mu\pi/2) + \frac{\rho g}{\mu}\sqrt{\frac{8\zeta}{\zeta^2+1}} \sin\!\big(\mu (\pi/2-\theta_0)\big) \cos(\mu \theta_0)=0. \label{eq:app:unequal-masses-odd-mu}
\end{equation}
For even-parity solutions, the equation is
\begin{equation}
\sin(\mu\pi/2) + \frac{\rho g}{\mu}\sqrt{\frac{8\zeta}{\zeta^2+1}} \sin\!\big(\mu (\pi/2-\theta_0)\big) \sin(\mu \theta_0)=0. \label{eq:app:unequal-masses-even-mu}
\end{equation}
Once $\mu$ is found, the wave function is also known cf.\ Eq. \eqref{eq:three-particles:wave-function} and the
energy is given by Eq. \eqref{eq:three-particles:harmonic-energy} \cite{loft2015a}.

The wave function in the infinite-interaction limit now vanishes along the delta-boundary lines $\phi = \pm\theta_0, \pm\pi/2, \pi\pm\theta_0$.
Only when $M=m$, is the wave function non-zero in all six regions separated by the delta-boundary lines.
In order to illustrate this, we look at two explicit examples where $M/m=1/2$ and $M/m=2$, respectively.
For the case of $M/m=1/2$, we have $\theta_0 \simeq 0.421$ (or $24.1^{\circ}$), and the lowest-energy solutions of Eqs. \eqref{eq:app:unequal-masses-odd-mu}--\eqref{eq:app:unequal-masses-even-mu} at $1/g=0$ have $\mu \simeq 0.731$ (both odd and even) while the second-lowest has $\mu \simeq 3.735$ (odd).
The angular part of the wave function for the ground state and the first excited state in the infinite-interaction limit are
\begin{equation}
\Phi_{\mu}^{(\mp)}(\phi) = \begin{cases}
\mp \sin(\mu(\phi+\theta_0)), & \text{for } \phi\in [-\frac{\pi}{2},-\theta_0] \\
0, & \text{for } \phi\in [-\theta_0,\theta_0] \\
\sin(\mu(\phi-\theta_0)), & \text{for } \phi\in [\theta_0,\frac{\pi}{2}],
\end{cases} \label{eq:app:angular-wave-mass-ratio-1-2}
\end{equation}
where $\mp$ is the parity of the state.
The angular part for the non-degenerate second excited state is
\begin{equation}
\Phi_{\mu}^{(-)}(\phi) = \begin{cases}
0, & \text{for } \phi\in [-\frac{\pi}{2},-\theta_0] \\
\cos(\mu\phi), & \text{for } \phi\in [-\theta_0,\theta_0] \\
0, & \text{for } \phi\in [\theta_0,\frac{\pi}{2}].
\end{cases} \label{eq:app:angular-wave-mass-ratio-2-1}
\end{equation}

When $M/m=2$, the roles are reversed.
The ground state is now non-degenerate with $\mu \simeq 2.552$ at $g=\infty$ and $\theta_0 \simeq 0.615$ (or $35.2^{\circ}$).
In addition, the wave function has the same form as Eq.~\eqref{eq:app:angular-wave-mass-ratio-2-1}, only with different $\theta_0$ and $\mu$.
For the first and second excited states, the wave function has the form of Eq.~\eqref{eq:app:angular-wave-mass-ratio-1-2}.

One might think that there would be some continuous crossover from $M<m$ to $M=m$ and then to $M>m$, but this is not the case.
Indeed, $M=m$ is a singular case where the wave function is non-zero in all six regions.


\section*{Acknowledgements}

Part of this work is based on the Bachelor thesis of M.E.S.A.
The authors thank C. Forss{\'e}n, D.~V. Fedorov, and A.~S. Jensen for discussions, as well as C. Forss{\'e}n and X. Cui for feedback on the 
manuscript. We thank the experimental team of the  S. Jochim group for extended discussions and for sharing their 
results.

This work was funded by the Danish Council for Independent Research DFF Natural Sciences and the DFF Sapere Aude program.

\section*{Author contributions statement}

A.G.V., E.J.L. and N.T.Z. devised the project. M.E.S.A. and A.S.D. developed the 
formalism under the supervision of A.G.V. and N.T.Z. The numerical calculations were carried out by A.S.D. and M.E.S.A. with assitance from A.G.V. and E.J.L. The initial draft of the paper was written by 
M.E.S.A., A.S.D., and N.T.Z. All authors contributed to the revisions that led to the final version.

\section*{Additional information}

The authors declare that they have no competing financial interests.


\section*{Supplementary Materials}

\subsection*{Stationary points of trial state energy functional}
\label{app:method:trial-state-energy-stationary-points}

For notational convenience, we define the function
\begin{equation}
E'(\alpha_0,\alpha_\infty) \equiv E - E_0 = \frac{\langle{\gamma_0|V|\gamma_0}\rangle\alpha_0^2 + \Delta E\alpha_\infty^2}{\alpha_0^2 + \alpha_\infty^2 + 2\langle{\gamma_0|\gamma_\infty}\rangle\alpha_0\alpha_\infty}.
\end{equation}
Stationary points of $E'$ are found where $(\partial E'/\partial\alpha_0,\partial E'/\partial\alpha_\infty)=(0,0)$. This gives the system of equations
\begin{equation}
\begin{bmatrix}
\langle{\gamma_0|V|\gamma_0}\rangle - E' & -\langle{\gamma_0|\gamma_\infty}\rangle E' \\
-\langle{\gamma_0|\gamma_\infty}\rangle E' & \Delta E-E'
\end{bmatrix} 
\left[\begin{matrix}
\alpha_0\\
\alpha_\infty
\end{matrix}\right]=0.
\label{eq:method:energy-functional-stationary-points-matrix}
\end{equation}

For non-trivial solutions, the determinant of the above coefficient matrix must be zero. This yields the quadratic equation
\begin{equation}
0 = \left(1-\langle{\gamma_0|\gamma_\infty}\rangle^2\right) E'^2 - \left(\langle{\gamma_0|V|\gamma_0}\rangle+\Delta E\right)E' + \langle{\gamma_0|V|\gamma_0}\rangle\Delta E
\end{equation}
from which we arrive at Eq. (7) of the Main Text.

If a pair, $(\alpha_0,\alpha_\infty)$, realizes a stationary point of $E'$, it solves Eq. \eqref{eq:method:energy-functional-stationary-points-matrix}.
This condition can be reduced to the relation
\begin{equation}
\alpha_0 = \frac{1}{\langle{\gamma_0|\gamma_\infty}\rangle} \frac{E_\infty-E}{E-E_0}\alpha_\infty.
\end{equation}
Substituting Eq. (7) of the Main Text for $E$, this gives Eq. (6) of the Main Text, or equivalently, that $\alpha_0/\alpha_\infty$ solves the quadratic equation
\begin{equation}
\langle{\gamma_0|V|\gamma_0}\rangle\langle{\gamma_0|\gamma_\infty}\rangle x^2 + (\langle{\gamma_0|V|\gamma_0}\rangle-\Delta E) x - \Delta E\langle{\gamma_0|\gamma_\infty}\rangle = 0.
\end{equation}

\subsection*{More on the two-particle system}

\subsubsection*{Wave functions}
\begin{figure}
\begin{center}
\includegraphics{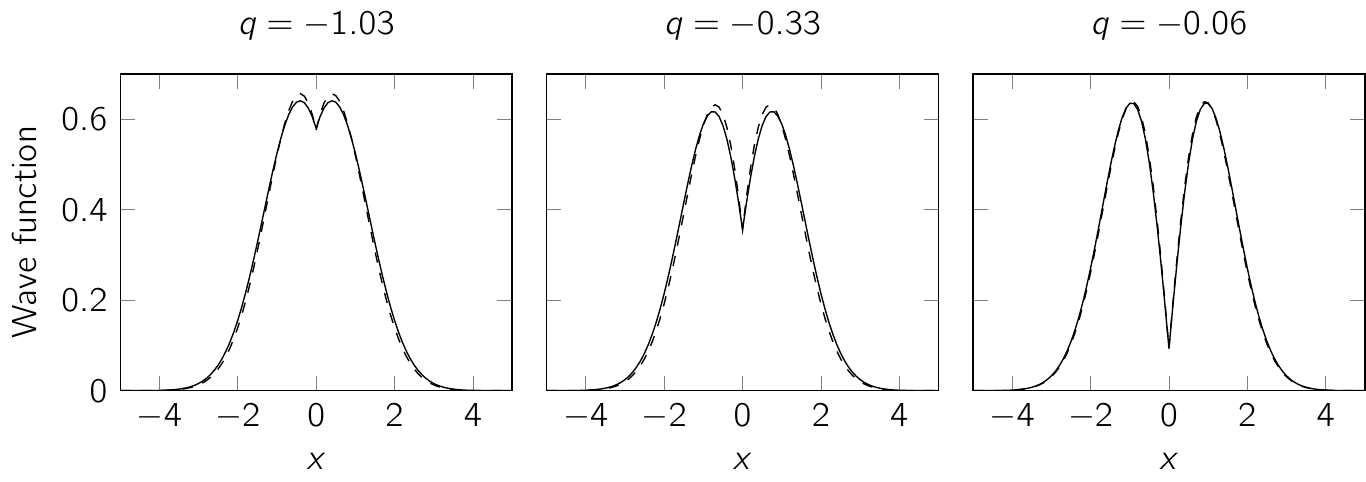}
\caption{{\bf Wave function of the two-particle system.}  The ground-state wave function of the relative motion of the two particles for different values of $q=-1/g<0$ according to the exact result (dashed) and the interpolatory ansatz (solid), respectively. The wave functions have been normalised.}
\label{fig:two-particles-ground-state-wave-function}
\end{center}
\end{figure}

As is shown in Fig.\ \ref{fig:two-particles-ground-state-wave-function}, the wave function of the ground state in the $q<0$ region as calculated with our ansatz is very similar to the exact one\cite{busch1998}.

\subsubsection*{Including a deeply bound state}
For attractive interactions, a deeply bound molecular state exists for which the energy diverges as $q$ approaches zero from the positive side. This state corresponds to the bound state of a delta potential well.

We may naively attempt to approximate the deeply bound state with the same trial state used for the ground state for $q<0$; that is, Eq. (4) of the Main Text with $(n_1,n_2)=(0,1)$.
The optimised energy of this state, however, does not diverge fast enough as $g \to-\infty$. In the following, we will present a revised ansatz for the deeply bound state. Note that 
this is in order to improve our accuracy in the strong bound regime (large negative energy). For weaker bound states (and for repulsive interaction) the interpolatory ansatz is very accurate without additional states.

For strong attractive interactions, the wave function of the deeply bound state is densely concentrated around $x=0$ and the length scale of the harmonic trap is very large compared to the extend of the wave function. The harmonic trap is thus neglectable when $g \ll 0$, and the wave function approaches that of a delta potential well, that is,
\begin{equation}
\langle{x|\gamma_\delta}\rangle = \sqrt{\frac{|g|}{2}}\,e^{-|g x|/\sqrt{2}},
\end{equation}
where we denote the ground state of the delta potential well by $|{\gamma_\delta}\rangle$ and its corresponding energy by $E = -\frac{1}{4}g^2$.

In this light, we may extend the ansatz with the additional state $|{\gamma_\delta}\rangle$:
\begin{equation}
|{\gamma}\rangle = \alpha_0|{\gamma_0}\rangle + \alpha_\infty|{\gamma_\infty}\rangle + \alpha_\delta|{\gamma_\delta}\rangle, \label{eq:two-particles:deeply-bound:extended-ansatz}
\end{equation}
where $|{\gamma_0}\rangle = |{n_1 = 0}\rangle$ and $|{\gamma_\infty}\rangle$ is given by Eq. (21) of the Main Text with $n_2=1$.
One might be tempted to leave out the state $|{\gamma_\infty}\rangle$, but this state is in fact required if the trial state is to approximate the ground state not only in the limits $g \simeq 0$ and $g \ll 0$, but also in-between.

\begin{figure}
\centering
\includegraphics{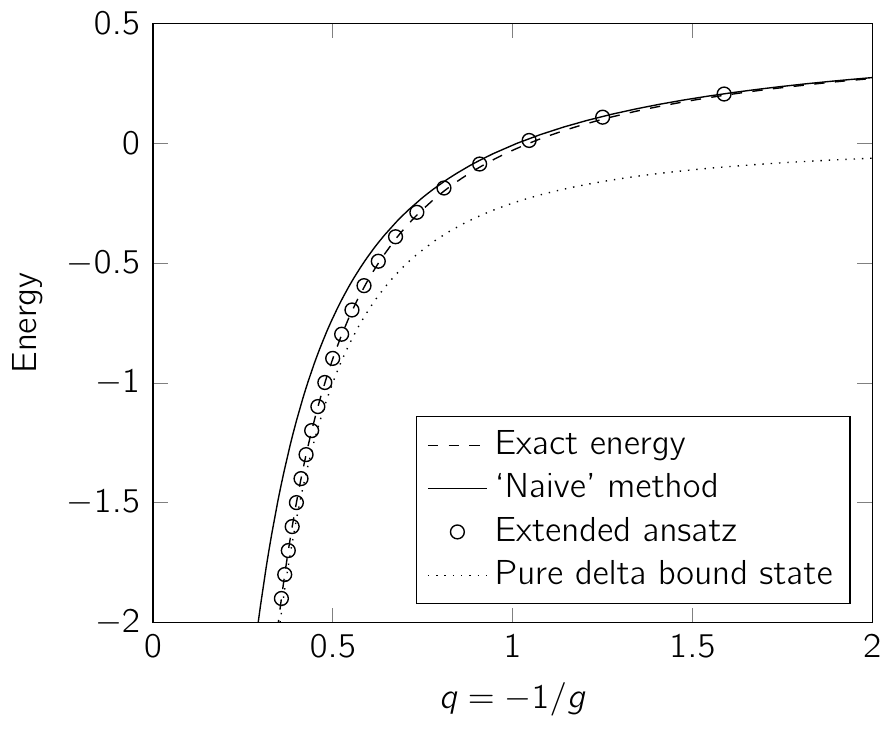}
\caption{{\bf Energy of deeply bound state.}  Energy of the two-particle ground state for attractive interactions according to the exact result (dashed), the `naive' interpolatory ansatz (solid) and the extended ansatz of Eq. \eqref{eq:two-particles:deeply-bound:extended-ansatz} (circles), respectively. For comparison, the ground-state energy of a system without the harmonic trap is also plotted (dotted).}
\label{fig:two-particles-delta-bound-state-energy-spectrum}
\end{figure}

\begin{figure}
\begin{center}
\includegraphics{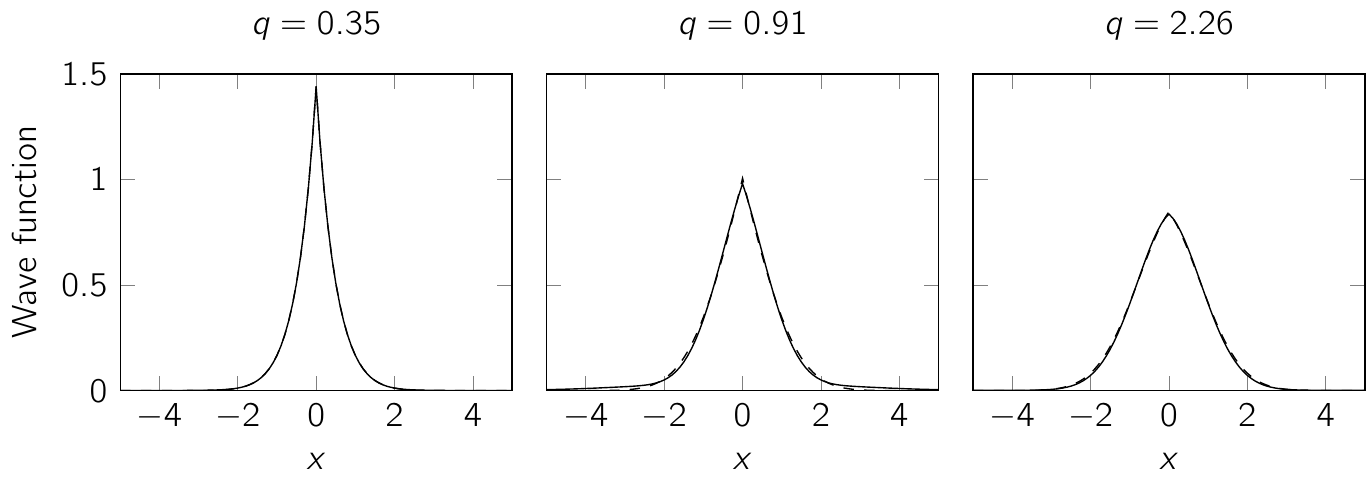}
\caption{{\bf Wave function of deeply bound state.}  The two-particle ground-state wave function for different values of $q=-1/g>0$ according to the exact result (dashed) and the extended ansatz of Eq. \eqref{eq:two-particles:deeply-bound:extended-ansatz} (solid), respectively. The wave functions have been normalised.}
\label{fig:two-particles-delta-bound-state-wave-function}
\end{center}
\end{figure}

The extended ansatz gives rise to a cubic equation containing error functions, and we have solved this numerically.
The resulting energy is shown in Fig.\ \ref{fig:two-particles-delta-bound-state-energy-spectrum} and is in much better agreement with the exact energy than the original ansatz. The wave function of the extended ansatz is also very close to the exact wave function. This is evident from Fig.\ \ref{fig:two-particles-delta-bound-state-wave-function}, comparing the two wave functions for three values of $q$.

\subsection*{Details of the impurity system}
\label{app:polaron:details}

Denote the wave function of the $n$'th excited state of the single-particle harmonic oscillator $\psi_n$ and define the Slater determinant \cite{girardeau}
\begin{align}
F(\psi_0,\dotsc,\psi_{N-1}; x_1,\dotsc,x_N) &\equiv \frac{1}{\sqrt{N!}} \det\!\begin{bmatrix}
\psi_0(x_1) & \psi_0(x_2) & \dots & \psi_0(x_N) \\
\psi_1(x_1) & \psi_1(x_2) & \dots & \psi_1(x_N) \\
\vdots & \vdots & \ddots & \vdots \\
\psi_{N-1}(x_1) & \psi_{N-1}(x_2) & \dots & \psi_{N-1}(x_N)
\end{bmatrix} \\
&= \left(\frac{2^{N-1}}{\pi}\right)^{N/4} \frac{1}{\sqrt{N!}} \left(\prod_{n=0}^{N-1} \frac{1}{\sqrt{n!}}\right) e^{-\mathbf{x}^2/2} \prod_{1\le j<k\le N} (x_k-x_j).
\end{align}

The ground state in the non-interacting limit has energy $E_0 = (N_\uparrow^2 + 1)/{2}$, and with the notation defined above,
\begin{equation}
\langle{\mathbf{x}|\gamma_0}\rangle = \psi_0(x_1) F(\psi_0,\dotsc,\psi_{N-2}; x_2,\dotsc,x_N).
\end{equation}
Meanwhile, the completely antisymmetric state has wave function 
\begin{equation}
\langle{\mathbf{x}|\gamma_{\mathrm{A}}}\rangle = F(\psi_0,\dotsc,\psi_{N-1}; x_1,\dotsc,x_N)
\end{equation}
and energy $E_\infty = N^2/2$.

The interaction energy in the non-interacting state is
\begin{equation}
\langle{\gamma_0|V|\gamma_0}\rangle = (N-1) g \int_{-\infty}^{\infty}\mathrm{d}{x_1} \int_{-\infty}^{\infty}\mathrm{d}{x_3} \int_{-\infty}^{\infty}\mathrm{d}{x_4}\, \dotsi \int_{-\infty}^{\infty}\mathrm{d}{x_N}\, \left.\langle{\mathbf{x}|\gamma_0}\rangle^2\right|_{x_2=x_1}.
\end{equation}
For the ground state, this reduces to \cite{gharashi}
\begin{equation}
\langle{\gamma_0|V|\gamma_0}\rangle= \frac{\sqrt{2}}{\pi (N - 2)!} \Gamma(N - 1/2) g.
\end{equation}

Using Eq. (16) of the Main Text, the $\langle{\gamma_0|\gamma_\infty}\rangle$ may be calculated through
\begin{equation}
\langle{\gamma_0|\gamma_\mathrm{A}}\rangle_n = \binom{N-1}{n-1} \int_{-\infty}^{\infty}\mathrm{d}{x_1} \int_{-\infty}^{x_1}\mathrm{d}{x_2} \,\dotsi \int_{-\infty}^{x_1}\mathrm{d}{x_{n}} \int_{x_1}^{\infty}\mathrm{d}{x_{n+1}}\, \dotsi \int_{x_1}^{\infty}\mathrm{d}{x_N}\, \langle{\gamma_0|\mathbf{x}}\rangle\langle{\mathbf{x}|\gamma_\mathrm{A}}\rangle. \label{eq:polaron:partial-overlap}
\end{equation}

\subsubsection*{Slope of energy curve}

The slope of the ground-state energy curve at $g\to\infty$ is the greatest eigenvalue of the matrix \cite{volosniev2014}
\begin{equation}
\mathbf{A} = \begin{bmatrix}
\alpha_1 & -\alpha_1 & 0 & 0 & \dots & 0 & 0 & 0 \\
-\alpha_1 & \alpha_1 + \alpha_2 & -\alpha_2 & 0 & \dots & 0 & 0 & 0 \\
0 & -\alpha_2 & \alpha_2 + \alpha_3 & -\alpha_3 & \dots & 0 & 0 & 0 \\
\vdots & \vdots & \vdots & \vdots & \ddots & \vdots & \vdots & \vdots \\
0 & 0 & 0 & 0 & \dots & -\alpha_2 & \alpha_1 + \alpha_2 & -\alpha_1 \\
0 & 0 & 0 & 0 & \dots & 0 & -\alpha_1 & \alpha_1
\end{bmatrix}
\end{equation}
where
\begin{equation}
\alpha_n = \frac{N!}{(n-1)!(N-1-n)!} \int_{-\infty}^{\infty}\mathrm{d}{x_1} \int_{-\infty}^{x_1}\mathrm{d}{x_3} \,\dotsi \int_{-\infty}^{x_1}\mathrm{d}{x_{n+1}} \int_{x_1}^{\infty}\mathrm{d}{x_{n+2}}\, \dotsi \int_{x_1}^{\infty}\mathrm{d}{x_N} \left(\left.\frac{\langle{\mathbf{x}|\gamma_\mathrm{A}}\rangle}{x_1-x_2}\right|_{x_2=x_1}\right)^2. \label{eq:polaron:alpha-coefficients}
\end{equation}
The coefficients, $a_n$, in Eq. (16) of the Main Text are the components of the eigenvector of $\mathbf{A}$ corresponding to the greatest eigenvalue.
Note that $\alpha_n = \alpha_{N+1-n}$, which is the origin of the bisymmetry of $\mathbf{A}$.

\subsubsection*{Error in energy of interpolatory ansatz}

\begin{figure}
\centering
\includegraphics{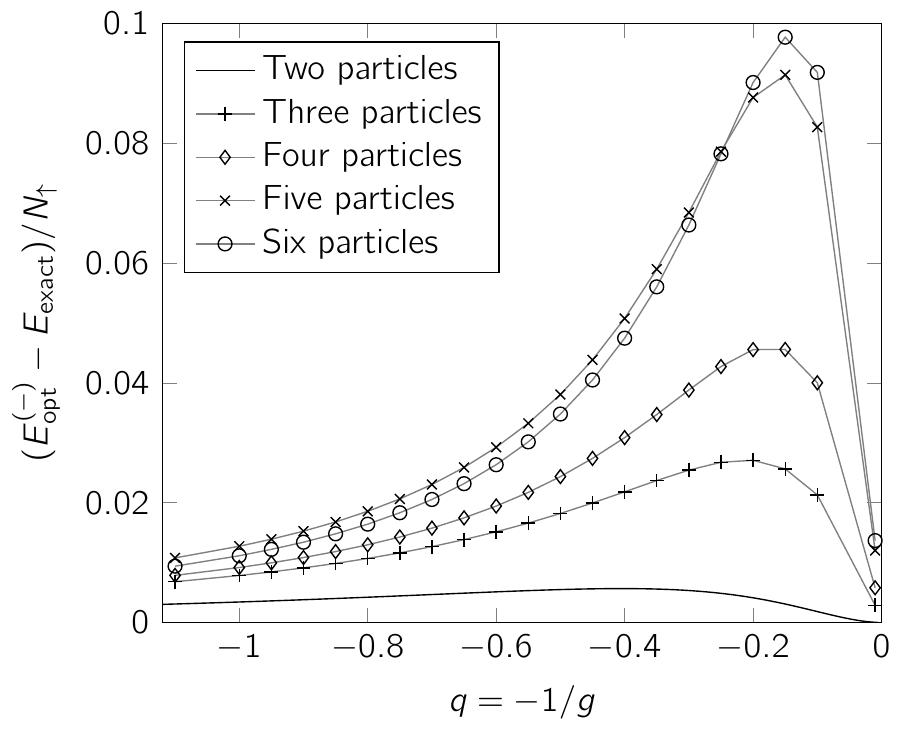}
\caption{{\bf Accuracy of the interpolatory ansatz applied to an impurity system.}  Error in energy according to the unmodified interpolatory ansatz compared to exact numerical results for impurity systems of $N=2-6$ particles. (For $N=2$ it is compared to the exact analytical solution.) For $N\ge 3$, the gray lines between the points are a mere guide to the eye.}
\label{fig:scaled-energy-error}
\end{figure}

Figure \ref{fig:scaled-energy-error} shows the error in the energy of the unmodifed interpolatory ansatz compared to exact numerical methods.

\subsubsection*{Optimum wave-function overlap}
\label{app:polaron:optimum-wave-function-overlap}

By Eq. (16) of the Main Text, the squared overlap is given by
\begin{equation}
\langle\gamma_0|\gamma_\infty(\mathbf{a})\rangle^2 = \frac{N}{\sum_{k=1}^N a_k^2} \left(\sum_{n=1}^N a_n \langle{\gamma_0|\gamma_\mathrm{A}}\rangle_n\right)^2. \label{eq:squared-overlap-polaron-states}
\end{equation}
Differentiating this with respect to a coefficient, $a_m$, and setting the result equal to zero, we get
\begin{equation}
0 = N \langle{\gamma_0|\gamma_\mathrm{A}}\rangle_m \sum_{n=1}^N a_n \langle{\gamma_0|\gamma_\mathrm{A}}\rangle_n - a_m \langle\gamma_0|\gamma_\infty(\mathbf{a}_\mathrm{max})\rangle^2,
\end{equation}
which is valid for $m=1,\dotsc,N$.
Thus, we arrive at the matrix equation
\begin{equation}
\left(N \Big[\langle{\gamma_0|\gamma_\mathrm{A}}\rangle_i \langle{\gamma_0|\gamma_\mathrm{A}}\rangle_j\Big]_{ij} - \langle\gamma_0|\gamma_\infty(\mathbf{a}_\mathrm{max})\rangle^2\right) \mathbf{a} = 0.
\end{equation}

This means that $\langle\gamma_0|\gamma_\infty(\mathbf{a}_\mathrm{max})\rangle^2$ and $\mathbf{a}_\mathrm{max}$ is an eigenvalue and a corresponding eigenvector of the matrix
\begin{equation}
N \Big[\langle{\gamma_0|\gamma_\mathrm{A}}\rangle_i \langle{\gamma_0|\gamma_\mathrm{A}}\rangle_j\Big]_{ij} =
N \begin{bmatrix}
\langle{\gamma_0|\gamma_\mathrm{A}}\rangle_1 \\ \langle{\gamma_0|\gamma_\mathrm{A}}\rangle_2 \\ \vdots \\ \langle{\gamma_0|\gamma_\mathrm{A}}\rangle_N
\end{bmatrix} \begin{bmatrix}
\langle{\gamma_0|\gamma_\mathrm{A}}\rangle_1 & \langle{\gamma_0|\gamma_\mathrm{A}}\rangle_2 & \dots & \langle{\gamma_0|\gamma_\mathrm{A}}\rangle_N
\end{bmatrix}.
\end{equation}
It is clear that the only non-zero eigenvalue belongs to the eigenvector $\mathbf{a}_\mathrm{max} = (\langle{\gamma_0|\gamma_\mathrm{A}}\rangle_1, \dotsc, \langle{\gamma_0|\gamma_\mathrm{A}}\rangle_N)$ and is as given in Eq. (17) of the Main Text; for there exist $N-1$ vectors orthogonal to $\mathbf{a}_\mathrm{max}$, each being an eigenvector with eigenvalue $0$.

\subsection*{Numerical methods}
\label{app:numerical-methods}

\subsubsection*{Effective interaction approach}

We consider a two-component system with $N_A$ particles in one component and $N_B$ particles in another, the so-called $N_A+N_B$ system.
Intra-species interactions are neglected, and all the particles are assumed to have the same mass, $m$, and trapping frequency, $\omega$.
The general Hamiltonian of the system can then be written as:
\begin{equation}
\mathcal{H}=\sum_{i=1}^{N_A}\left(\frac{p_{A,i}^2}{2m}+\frac{m\omega^2}{2} q_{A,i}^2\right)+\sum_{i=1}^{N_B}\left(\frac{p_{B,i}^2}{2m}+\frac{m\omega^2}{2} q_{B,i}^2\right)+\sum_{i_A=0}^{N_A}\sum_{i_B=0}^{N_B}V_{i_A,i_B}
\label{eq:manyhamiltonian}
\end{equation}
where $V_{i_A,i_B} = g\delta(q_{i_A}-q_{i_B})$ are the interaction terms ($g$ being the interaction strength), and the first two parentheses are the non-interacting part of the Hamiltonian; call it $H_0$.
$p_{k,i}$ and $q_{k,i}$ are the momentum and coordinate operators, respectively, for particle $i$ in subsystem $k\in\{A,B\}$.
They each operate in their own subspace, so $p_{A,i}=p_{i}\otimes 1$ and $p_{B,i}=1 \otimes p_{i}$.

The total many-body basis state is a tensor product of many-body states from each species.
We refer to the states from each subsystem as few-body states and to states describing the full system as many-body states.
In each subsystem we have identical fermions and therefore we need a totally antisymmetric few-body state, that is, a state that is antisymmetric under the exchange of \emph{any two} particles:
\begin{equation}
|(m_1 m_2 \cdots m_N)\rangle \equiv \frac{1}{\sqrt{{N!}}} \sum_{\sigma\in S_N} |m_{\sigma(1)}\rangle |m_{\sigma(2)}\rangle \cdots |m_{\sigma(N)}\rangle,
\end{equation}
where we choose $m_1> m_2> \dots > m_N$ by convention and $S_N$ is the symmetric group of order $N$.
The $|m_i\rangle$ represents a single-particle state equivalent to a harmonic oscillator eigenstate corresponding to eigenvalue $m_i$ with respect to the number operator $a_i^\dagger a_i$.
These single-particle states are convenient to use since they are eigenstates of $H_0$.
The complete basis for the full system can thus be written as
\begin{equation}
|\Psi\rangle=|(m_1 m_2 \cdots m_{N_A})\rangle\otimes|(k_1 k_2 \cdots k_{N_B})\rangle.
\end{equation}
The corresponding eigenvalue to $H_0$ is $E=\hbar\omega(\frac{N_A+N_B}{2}+m_1+\dots+m_{N_A}+k_1+\dots+k_{N_B})$.

One of the nice properties of the Hamiltonian is that its kinetic energy and harmonic trap operators are one-particle operators, and only the interaction operator couples the particles.
This means that the overall matrix is actually a sparse matrix.
By construction, the contribution of the kinetic energy and harmonic trap terms are trivial.
However, the interaction part is given as
\begin{equation}
M=\mathrm{sgn}[\sigma\tau\sigma'\tau'] V_{n_{\sigma(1)},h_{\tau(1)},n'_{\sigma'(1)},h'_{\tau'(1)}},
\label{eq:matrixelementsreduced}
\end{equation}
where $V_{a,b,c,d}\equiv\langle a,b|V|c,d\rangle$ is the two-body subspace matrix element and $\mathrm{sgn}$ is the sign function, which comes from how many times the states are swapped with each other.
In addition, we are only interested in the intrinsic dynamics of states, therefore we use a Lawson projection term \cite{lawson1974} to push away the many-body solutions corresponding to excitations of the center of mass.

An effective two-body interaction is considered instead of the bare zero-range interaction. The advantage of this effective interaction is that it converges rapidly as a function of model space size. This has been utilized to address cold atomic gases in recent papers \cite{christensson2009,rotureau2013}. It is constructed in a truncated two-body space, $P$, defined as the set of two-body relative harmonic oscillator states whose radial quantum numbers are smaller than a cutoff, $n_\mathrm{max}$, and it is designed such that its solutions correspond to the two-body energies that are given by the Busch formula~\cite{busch1998}. The unitary transformation of the constructed two-body effective Hamiltonian is given as~\cite{lindgren2014}
\begin{equation}
H_p^\mathrm{eff}=\frac{U_{PP}^\dagger}{\sqrt{(U_{PP}^\dagger U_{PP})}}E_{PP}^{(2)}\frac{U_{PP}^\dagger}{\sqrt{(U_{PP}^\dagger U_{PP})}}
\end{equation}
where $E_{PP}^{(2)}$ is the diagonal matrix with eigenvalues from the $P$-space and $U_{PP}$ is the matrix whose rows are formed by the corresponding eigenvectors. In the limit of infinite model space, $n_\mathrm{max}\to\infty$, the unitary transformation approaches the exact bare Hamiltonian results. However, the convergence to the exact limit is a lot quicker than expected for small systems. For example, $n_\mathrm{max}=20$ for the $2+1$ system is more than enough to obtain results with a precision of 3 decimals and only a few minutes of calculation time.

\subsubsection*{Correlated Gaussian approach}
Here we present the details of the correlated Gaussian approach that we have 
employed for the mass-imbalanced case. For further details on this method see Ref.~\cite{mitroy2013}.
We consider a general Hamiltonian given as
\begin{equation}
\mathcal{H}=\sum_{i=1}^{N_\mathrm{tot}}\left(\frac{p_i^2}{2m_i}+V_\mathrm{ext}(x_i)\right)+\sum_{i<j}V(x_i,x_j),
\end{equation}
where $x_i$ is the coordinate of the $i$'th particle, $V_\mathrm{ext}(x_i)$ is the external confinement and $V(x_i,x_j)$ is the 2-body interaction potential. In our case $V_\mathrm{ext}(x_i)=\frac{1}{2}m\omega^2x_i^2$ and $V(x_i,x_j)=g\delta(x_i-x_j)$, with $g$ being the interaction strength. Please note that the specified potentials are not crucial for the method and any other system with vanishing 2-body potential for larger separation and any bounded external potential could easily do. An upper limit for the bound ground state can be found variationally based on the functional
\begin{equation}
E_\text{upper}[f]=\frac{\langle f|H|f\rangle}{\langle f|f\rangle},
\end{equation}
where $|f\rangle$ is a normalizable and differentiable function built as a linear combination of states from a basis $\{\phi_k\}$: $|f\rangle=\sum_{k=1}^l c_k\phi_k$, where $l$ is a computationally accessible number that is set to reach a given precision. We use a basis in the form of Gaussian functions,
\begin{equation}
\phi_k=e^{-(x_j-s_j^k)A_{jf}^k(x_f-s_f^k)}\equiv e^{-(\mathbf{x}-\mathbf{s})^T\mathbf{A}(\mathbf{x}-\mathbf{s'})},
\end{equation}
where $\{x_j\}$ are the coordinates of the system while $s_j^k$ and $A_{jf}^k$ are numbers that characterize the basis elements. In order to ensure square-integrability we assume $A_{jf}^k$ is symmetric and positive-definite. Note that Einstein's repeated summation notation is used here.

Gaussian functions usually have an analytical expression when one wants to calculate for instance $\langle \phi_k|\phi_k\rangle$ or $\langle \phi_k|\frac{\partial^2}{\partial x^2}|\phi_k\rangle$, making it very fast to calculate such expressions numerically. These functions can also be transformed easily from one Jacobi coordinate set to another and even any desirable features such as symmetry can be implemented into the ansatz.

Our next step is to choose a subset of $l$ elements from a complete basis. This can be done in many ways, deterministically, randomly or a mix. In our calculations, we choose the first $k$ elements with $k<l$ stochastically. This subset is the starting point of the trial function. We generate $A_{jl}^k$ and $s_j^k$ randomly from an appropriate distribution (e.g. exponential or Laplacian) and then determine the upper bound for the ground state. We can choose to do this step several times, say $\alpha$ times, and then among these $\alpha$ times we choose the best trial function with the lowest trial energy, call it $k_\text{best}$. Then we can start to expand $k_\text{best}$ by adding some other elements randomly created, $k_\text{best}\mapsto k_\text{best}+k_\text{add}\leq l$ and by doing so, say $\beta$ times, and then again picking the lowest energy among the $\beta$ trials as our new candidate, we end up constructing a trial function that has an upper bound for the ground-state energy.

One should note, that in some situations where the functions do not decay fast enough at infinity or there are some delta functions, the number of elements in the finite basis has to be very large or go to infinity in order to describe the exact wave function everywhere. For the convergence and error of this method, see Ref. \cite{volosniev-thesis}. However, the method used here for our system converges relatively fast with a precision up to 4 decimals with the parameters $\alpha=500$ and $\beta=500$ and a calculation time of approximately 1 hour.

In order to illustrate the method, we look at our Hamiltonian for the mass-imbalanced $2+1$ system given in relative coordinates:
\begin{equation}
H_\text{rel}=T_\text{rel}+V_\text{ext,rel}+\sum_{i<j}V_{ij}
\end{equation}
where $T_\text{rel}=\frac{1}{2}(p_{x_1'}^2+p_{x_2'}^2)$, $V_\text{ext,rel}=\frac{1}{2}(x_1'^2+x_2'^2)$ and 
\begin{equation}
\sum_{i<j}V_{ij}=\frac{g}{\hbar\omega\sigma}\left[\frac{\mu_{23}}{\sqrt{\mu}}\delta(x_1')+\frac{\sqrt{\mu}}{\mu_{23}}\delta\!\left(\frac{\mu}{m_1}x_1'+x_2'\right)+\frac{\sqrt{\mu}}{\mu_{23}}\delta\!\left(-\frac{\mu}{m_2}x_1'+x_2'\right)\right],
\end{equation}
with $x_1'$, $x_2'$, $\sigma$, $\mu$ and $\mu_{23}$ defined as in the treatment of the mass-imbalanced system in the main text. We ignore the center-of-mass Hamiltonian because its solutions are already known.

With this Hamiltonian, we can calculate the following quantities:
\begin{equation}
N_{ij}  \equiv \langle\phi_i|\phi_j\rangle = \exp\!\left(-(\mathbf{s}^T\mathbf{A}\mathbf{s}
+\mathbf{s}'^T\mathbf{A}'\mathbf{s}')+\frac{1}{4}{\mathbf{v}^T
(\mathbf{B}^{-1})^T\mathbf{v}}\right) \frac{\pi^{3/2}}{\sqrt{\det(\mathbf{B})}}
\end{equation}
\begin{equation}
T_\text{rel}  = \frac{1}{2}\langle\phi_i|\mathbf{p}^T\mathbf{p}|\phi_j\rangle = \frac{1}{2}N_{ij} \left(\mathrm{Tr}(\mathbf{A}')-\mathrm{Tr}(\mathbf{A}'^2\mathbf{B}^{-1})-2\boldsymbol{\alpha}^T\mathbf{A}'^2
\boldsymbol{\alpha}+4\mathbf{s}'^T\mathbf{A}'^2\boldsymbol{\alpha}-2
\mathbf{s}'^T\mathbf{A}'^2\mathbf{s}'\right)
\end{equation}
\begin{equation}
V_\text{ext,rel} = \frac{1}{2}\langle\phi_i|\mathbf{x}_1'^T\mathbf{x}|\phi_j\rangle = \frac{1}{2}N_{ij} \left(\frac{1}{2}\mathrm{Tr}(\mathbf{B}^{-1})+\frac{1}{4}\mathbf{v}^T (\mathbf{B}^{-1})^2 \mathbf{v}\right)
\end{equation}
\begin{equation}
\langle\phi_i|\delta(x)|\phi_j\rangle=\frac{\sqrt{\det(\mathbf{B})}}{\pi^{3/2}}N_{ij} \exp\!\left({-\det(\mathbf{B}) \frac{\alpha_1^2}{B_{22}} }\right)
\end{equation}
\begin{align}
\langle\phi_i|\delta\!\left(\pm\frac{\mu}{m}x_1'+x_2'\right)|\phi_j\rangle &= \frac{\sqrt{\det(\mathbf{B})}}{\pi^{3/2}}N_{ij}\frac{m}{\mu}\frac{\sqrt{\pi}}{\sqrt{B_{22}+\frac{m^2}{\mu^2}B_{11}\mp\frac{m}{\mu}2B_{12}}} \exp\!\left({-\det(\mathbf{B}) \frac{(\alpha_1\pm\frac{m}{\mu}\alpha_2)^2}{B_{22}+\frac{m^2}{\mu^2}B_{11}\mp\frac{m}{\mu}2B_{12}}}\right),
\end{align}
where $\mathbf{B}=\mathbf{A}+\mathbf{A}'$, $\mathbf{v}=2(\mathbf{A}\mathbf{s}+\mathbf{A}'\mathbf{s}')$, $\boldsymbol{\alpha}=\frac{1}{2}\mathbf{B}^{-1}\mathbf{v}$ and $m_2=m_3=m$. The analytical expressions for each of the integrals make it easy to calculate the value of $\langle f|H|f\rangle$ fast numerically and then try this several times for some randomly generated ansatz. In this way we can find a close upper limit to the ground-state energy of our Hamiltonian.

\end{document}